\documentclass[a4paper,10pt]{article}
\usepackage[utf8]{inputenc}
\usepackage{amssymb}

\usepackage{booktabs}

\usepackage{dcolumn}
\usepackage{bm}
\usepackage{verbatim}  
\usepackage{relsize}
\usepackage{xspace}
\usepackage{fixltx2e} 
\usepackage{graphicx,amsmath,amsfonts,amscd,amssymb,url}

\usepackage[T1]{fontenc}
\usepackage[utf8]{inputenc}
\usepackage{authblk}

\newcommand{\qty}[2]{\ensuremath{#1\,\mathrm{#2}}}
\newcommand{\bstar}{\ensuremath{\beta^{*}}}
\newcommand{\bstarval}[1]{$\bstar = #1\,\mbox{m}$}
\newcommand{\murad}[1]{\qty{#1}{\mu rad}} 
\newcommand{\mum}[1]{\qty{#1}{\mu m}}  
\newcommand{\sigmatot}{\ensuremath{\sigma_{\mathrm{tot}}}}

\newcommand{\bs}{$\beta^*$\xspace}

\newcommand{\nb}{nb$^{-1}$\xspace}

\newcommand{\um}{$\mathrm{\mu m}$\xspace}
\newcommand{\urad}{$\mathrm{\mu rad}$\xspace}

\newcommand{\lum}[1]{\qty{#1\times 10^{27}}{cm^{-2}s^{-1}}}

\newcommand{\lumTN}[1]{\qty{#1\times 10^{29}}{cm^{-2}s^{-1}}}
\newcommand{\expfor}[2]{$#1\!\times\! 10^{#2}$}

\newcommand{\Lu}{\mathcal{L}}
\newcommand{\Lum}{\ensuremath{\mathcal{L}}}
\newcommand{\Tf}{T_\mathrm{f}}
\newcommand{\Tpibs}{$T_{p,\mathrm{IBS}}$\xspace}
\newcommand{\Txibs}{$T_{x,\mathrm{IBS}}$\xspace}
\newcommand{\Tyibs}{$T_{y,\mathrm{IBS}}$\xspace}
\newcommand{\Tuibs}{T_{u,\mathrm{IBS}}}
\newcommand{\Tradp}{$T_{p,\mathrm{rad}}$\xspace}
\newcommand{\Tradx}{$T_{x,\mathrm{rad}}$\xspace}
\newcommand{\Tradu}{T_{u,\mathrm{rad}}}
\newcommand{\Tcore}{$T_{xy,\mathrm{core}}$\xspace}
\newcommand{\Teff}{$T_{x,\mathrm{eff}}$\xspace}

\newcommand{\Tta}{T_\mathrm{ta}}
\newcommand{\Topt}{T_\mathrm{f,opt}}
\newcommand{\Trun}{T_\mathrm{run}}
\newcommand{\Lavg}{\Lu_\mathrm{avg}}
\newcommand{\Ltot}{\Lu_\mathrm{tot}}
\newcommand{\fs}[4]{\texttt{#1b\char`_#2\char`_#3\char`_#4}}\xspace

\newcommand{\exy}{\ensuremath{\epsilon_{xy}}}

\newcommand{\intlum}{$\int \Lu \,dt$\xspace}
\newcommand{\Tnc}{\ensuremath{T_\mathrm{nc}}}

\title{Performance and luminosity models for heavy-ion operation at the CERN Large Hadron Collider}
\author[1]{R. Bruce\thanks{roderik.bruce@cern.ch}}
\author[1,2]{M.A.~Jebramcik}
\author[1,3]{J.M.~Jowett}
\author[4]{T. Mertens}
\author[1]{M. Schaumann}
\affil[1]{CERN, Geneva, Switzerland}
\affil[2]{DESY, Hamburg, Germany}
\affil[3]{GSI Helmholtzzentrum für Schwerionenforschung, Darmstadt, Germany}
\affil[4]{Helmholtz-Zentrum Berlin fur Materialien und Energie, Berlin Germany}

\begin{document}

\maketitle

\begin{abstract}

A good understanding of the luminosity performance in a collider, as well as reliable tools to analyse, predict, and optimise the performance, are of great importance for the successful planning and execution of future runs. 
In this article, we present two different  models for the evolution of the beam parameters and the luminosity in heavy-ion colliders. 
The first, Collider Time Evolution (CTE) is a particle tracking code, while the second, the Multi-Bunch Simulation (MBS), is based on the numerical solution of ordinary differential equations for beam parameters. 
As a benchmark, we compare simulations and data for a large number of physics fills in the 2018 Pb-Pb run at the CERN Large Hadron Collider (LHC), finding excellent agreement for most parameters, both between the simulations and with the measured data. 
Both codes are then used independently to predict the performance in future heavy-ion operation, with both Pb-Pb and p-Pb collisions, at the LHC and its upgrade, the High-Luminosity LHC. 
The use of two independent codes based on different principles gives increased confidence in the results.

\end{abstract}

\section{Introduction}
\label{sec:intro}
After the centre-of-mass energy, perhaps the most important figure of merit of any particle collider is the time integral of the luminosity delivered to the particle or nuclear physics experiments. 
Generally, integrated luminosity is a measure of the data collected and defines the precision of measurements and the potential for new discoveries\footnote{In nucleus-nucleus (or ``heavy-ion'') colliders, the total hadronic event rate may be limited by detector capabilities, especially in ``minimum-bias'' data-taking where a large fraction of all events are recorded and the data accumulation rate is especially high.  
The average number of events per bunch crossing (``pile-up") may also need to be limited.  
To achieve this, in practice, the overlaps or the sizes of the colliding beams are adjusted to level the luminosity at the maximum acceptable value.}.   

The product of luminosity \Lum\ and the appropriate cross-section is the rate of events of a specific type. During a fill of a circular collider, the phase-space distributions and intensities of the beams evolve in time under  the combined influences of several inter-dependent physical processes, 
e.g., the beam-beam collisions (or luminosity losses), 
intra-beam scattering (IBS), and synchrotron radiation damping. 
As a consequence of the intensity losses,  \Lum\ eventually 
decreases with time.  
To maximise \intlum, non-collisional losses must be minimised and the beam sizes should be kept small. 
To understand, predict, and optimise the physics data collected, it is therefore very important to have reliable models of the quantitative time evolution of the beams and \Lum. 
In this article we present two such models: Collider Time Evolution (CTE), and the Multi-Bunch Simulation (MBS). 
They have been developed for heavy-ion operation at the CERN Large Hadron Collider (LHC)~\cite{lhcdesignV1}, although they are also applicable to other hadron  colliders. 

There have been numerous studies of the time evolution of beams and   collider luminosity in the past, e.g., those in Refs.~\cite{hubner85,morsch94,baltz96,wei90,gounder03,sidorin06,blaskiewicz-cool07,blaskiewicz08,bruce10prstabCTE,antoniou16_evian}, with similar applications to this article. 
Most of them rely on the numerical solution of systems of coupled ordinary differential equations (ODEs) that model the time-evolution of a few key beam parameters, most commonly the transverse beam emittances, the bunch lengths  and intensities. 
These models use additional assumptions on the beam distributions, which are typically assumed to be Gaussian in all three degrees of freedom. 
The beam distributions can also be tracked by solving   Fokker-Planck equations for distribution functions, as in Refs.~\cite{wei90,gounder03}. 
Past studies of the heavy-ion luminosity at the LHC were carried out in Ref.~\cite{epac2004} using ODEs, and in Ref.~\cite{bruce10prstabCTE}, where a  particle tracking simulation was used in addition to the ODE method. 
That study used an early version of the CTE code, which has since   been further developed and significantly improved. 

Since the LHC is used for the benchmark, we give first a brief introduction to heavy-ion operation of the LHC in Sec.~\ref{sec:LHC}. 
Then we present the simulation codes, CTE in Sec.~\ref{sec:cte} and MBS in Sec.~\ref{sec:mbs}, and  compare their results to data from a large number of fills in the 2018 LHC Pb-Pb run in Sec.~\ref{sec:2018}. 
Finally we use the simulations to predict the performance in future LHC heavy-ion runs in Sec.~\ref{sec:hl-lhc}.

\section{Heavy-ion operation of the LHC}\label{sec:LHC}

\begin{table*}[tbh]
  \centering
    \begin{tabular}{lrrr}     \hline        
                                   & LHC     & 2018          & Run 3 and \\  
                                   & design  &               & HL-LHC \\ \hline
Beam energy (\qty{Z}{TeV})                 & 7             & 6.37          & 7 \\
Total no.\ of bunches                  & 592           & 733           & 1240 \\  
Bunch spacing (ns)                  & 100           & 75            & 50   \\  
Bunch intensity ($10^7$ Pb ions)    & 7             & 21            & 18 \\  
Stored beam energy (MJ)             & 3.8           & 12.9          & 20.5\\  
Normalized transverse emittance (\um)& 1.5          & 2.3           & 1.65           \\  
Longitudinal emittance (eVs/charge) & 2.5           & 2.33          & 2.42           \\  
RMS energy spread ($10^{-4}$)       & 1.1           & 1.06          & 1.02            \\  
RMS bunch length (cm)               &  7.94         & 8.24          & 8.24          \\  
Number of colliding bunches (IP1/5) &$<592$            & 733         & 976--1240$^a$ \\  
Number of colliding bunches (IP2)   &592            & 702           & 976--1200$^a$ \\  
Number of colliding bunches (IP8)   & 0             & 468           & 0--716$^a$     \\  
\bstar\ at IP1/5 (m)                    & 0.55           & 0.5           & 0.5 \\  
\bstar\ at IP2 (m)                      & 0.5           & 0.5           & 0.5 \\  
\bstar\ at IP8 (m)                      & 10.0          & 1.5           & 1.5 \\  
half crossing angle, IP1/5 (\urad)        & 160            & 160            & 170 \\  
half crossing angle, IP2 (external,net) (\urad)&110,40     & 137,60       & 170,100 \\  
half crossing angle, IP8 (external,net) (\urad)&---        & 160          & -170,-305 \\  
Peak luminosity, IP1/2/5 ($10^{27}$~cm$^{-2}$s$^{-1}$)& 1.0 &   6.1 & --- \\  
Levelled luminosity, IP1/5 ($10^{27}$~cm$^{-2}$s$^{-1}$)& --- &  --- &  6.4 \\  
Levelled luminosity, IP2 ($10^{27}$~cm$^{-2}$s$^{-1}$)& --- &  1.0 &  6.4 \\  
Levelled luminosity, IP8 ($10^{27}$~cm$^{-2}$s$^{-1}$)& --- &  1.0 &  1.0 \\ \hline
\end{tabular}%
   \footnotesize{
\\ $^a$ The values give the range over the filling patterns considered in Ref.~\cite{bruce20_HL_ion_report}. \\
}
   
   \caption{\label{tab:lhc} Pb beam parameters at the start of collisions in the LHC, as foreseen in the LHC design report (two experiments illuminated)~\cite{lhcdesignV1}, as achieved in 2018~\cite{jowett19_evian,jowett19_ipac}, and as envisaged for future runs in Run~3 (presently planned from 2022) and in HL-LHC~\cite{jowett17_cham,bruce20_HL_ion_report}. The 2018 parameters refer to typical in the fills with 75~ns bunch spacing. The crossing angles refer to the vertical plane in IR1 and IR2 and to the horizontal plane in IR5 and IR8, although it should be noted that the IR1/5 crossing planes may be swapped for HL-LHC to follow the proton configuration. }%
\end{table*}

The LHC is a circular collider, designed to collide protons and nuclei at a beam energy of $7\,Z$~TeV;  so far up to $6.5\,Z$~TeV has been achieved\footnote{As usual,  $Z$ is the atomic number of the nuclei forming the beam.}. 
The LHC uses two counter-rotating beams, called B1 and B2, that collide at four interaction points (IPs), inside the  experiments ATLAS~\cite{atlas} at IP1, 
ALICE~\cite{alice04} at IP2, 
CMS~\cite{cms} at IP5, and LHCb~\cite{lhcb_Jinst} at IP8. 
Typically, the LHC has operated for about one month per year with heavy-ion beams, mainly fully stripped Pb nuclei, in addition to the main physics programme with proton-proton collisions. 
The initial aim was mainly to provide Pb-Pb collisions to ALICE, which is specialised in heavy-ion physics, but over time all the  LHC experiments have joined the heavy-ion  collision programme. 

Four Pb-Pb runs were carried out so far (in 2010, 2011, 2015, and 2018)~\cite{ipac11_jowett_fist_Pb_run,jowett16_ipac,jowett19_ipac}, resulting in  \intlum$\approx1.5$~\nb at ALICE, thus surpassing the initial target of 1~\nb for the first 10~years of operation~\cite{alice04} in spite of more experiments than initially foreseen sharing the luminosity and a lower-than-nominal beam energy (\qty{3.5\,Z}{TeV} in Run~1 (2010--2013)\footnote{We follow the usual convention at CERN of referring to annual operating periods as ``runs'' and multi-year operating periods between long shutdowns as (capitalised) ``Runs''.} 
and  \qty{6.37\,Z}{TeV} in Run~2 (2015--2018), as opposed to the nominal \qty{7\,Z}{TeV}). 
So far, ALICE needed to be levelled at \lum{1} in order not to exceed the event rate limit of the detector. 
The high-luminosity experiments ATLAS and CMS collected in total about 2.5~\nb in Run~1 and Run~2, since no luminosity levelling was needed, or  only done at a much higher value~\cite{jowett19_evian}. LHCb was the last experiment to join the heavy-ion programme, integrating about 0.25~\nb in Run~2. 

In addition to Pb-Pb collisions, a completely new mode of operation with proton-nucleus collisions was put in place~\cite{jowett06,jowett13_pPb,versteegen13,jowett17_ipac_pPb}. 
Runs with p-Pb, not foreseen at the LHC design stage, were carried out in 2012, 2013 and 2016. 
About 250~\nb were gathered in ATLAS and CMS, and 75~\nb in ALICE. 

The key parameters of the LHC in Pb-Pb mode are shown in Table~\ref{tab:lhc}, where we highlight first the design parameters~\cite{lhcdesignV1} and those achieved in  2018~\cite{jowett19_ipac}. 
Notably,  most of the LHC design parameters have been reached or surpassed;  in particular the 2018 luminosity was more than a factor~6 higher than the design value. 
This was achieved mainly thanks to a factor~3 larger bunch population from the injectors and a larger number of bunches. 
A new filling pattern with a shorter bunch spacing of 75~ns, compared to the previous 100~ns spacing, was adopted half-way through the 2018 run. 

After the very successful first two Runs, the heavy-ion programme is scheduled to continue in the future, typically with short runs of about 1~month per year of Pb-Pb or p-Pb collisions. 
The projected future parameters~\cite{bruce20_HL_ion_report} are shown in Table~\ref{tab:lhc}. 
The physics goals and luminosity targets of the experiments are specified in Ref.~\cite{YR_WG5_2018,Bruce_ion_physics_BSM_2020}. 
The LHC performance in these future runs will benefit from upgrades of the LHC injectors~\cite{liu_TDR_ions} and the high-luminosity LHC (HL-LHC)~\cite{hl-lhc-tech-design}. 
Thanks to the injector upgrades, an even shorter bunch spacing of 50~ns will be achievable in the future. 
To safely store the higher beam intensity without an increase of spurious beam dumps and quenches, it is planned to upgrade the LHC collimation system~\cite{assmann05chamonix,assmann06,bruce14_PRSTAB_sixtr,bruce14ipac_DS_coll}. 
Upgrades of the ALICE detectors~\cite{ALICE_LOI_2014} will allow levelling at a  significantly higher hadronic event rate of 50~kHz corresponding to a luminosity of $\Lum=\lum{6.4}$. 
The feasibility of this luminosity also relies on the installation of new collimators~\cite{hl-lhc-tech-design} to intercept collision products that risk to quench impacted magnets, in particular from bound-free pair production (BFPP) and to a lesser extent from electromagnetic dissociation~\cite{klein01,epac2004,prl07,prstabBFPP09,jowett16_ipac_bfpp,schaumann20_PRAB_BFPP}. 
The luminosity in LHCb will still be limited by the BFPP losses, since no new collimators are foreseen to be installed in IR8. 

All upgrades of the injectors, collider, and experiments relevant to the heavy-ion programme are planned to be implemented for the machine startup after long shutdown 2 (LS2), presently foreseen for 2022, making the full heavy-ion ``HL-LHC'' performance available~\cite{bruce20_HL_ion_report}. A proposal for future running with lighter nuclei has also been put forward~\cite{YR_WG5_2018}. 

\section{Collider Time Evolution (CTE)}
\label{sec:cte}

\subsection{General features}

The CTE code, initially presented and benchmarked with data from the Relativistic Heavy-Ion Collider (RHIC)   in Ref.~\cite{bruce10prstabCTE}, and further developed in Refs.~\cite{Mertens-thesis,michaela_thesis} and for this article, tracks bunches of macro particles using a one-turn map. Typically we simulate $2\times10^4$ macro particles per bunch, which are tracked in a 6D phase space. 
On every turn, various physical effects are sequentially applied to the bunches and the particle coordinates are updated accordingly. 

The physical effects in CTE, which can be turned on individually in any combination, include betatron motion, longitudinal motion, beam-beam collisions with optional luminosity levelling, IBS, radiation damping and quantum excitation, machine aperture, extra generic losses modelled through a non-collisional lifetime, generic emittance blowup modelled through a rise time, and stochastic cooling\footnote{There is no stochastic cooling  in the LHC and it was not used in all simulations shown in this article.}.  

The user can choose to simulate a smaller number of turns than the real machine turns to increase the computational speed. 
The strengths of the physical processes, e.g., collision probabilities or amplitude changes from IBS or radiation damping, are then scaled up according to the real time of the simulated fill. For the results shown in this article we typically substitute $4\times10^8$ machine turns, corresponding to about 10~h of LHC operation, by $2\times10^4$ simulation turns. 

The strengths of the physical processes are scaled further to account for the smaller number of macro particles in the simulated bunches than real particles in the LHC bunches. Mostly we use $2\times10^4$ macro particles per bunch. This gives a fast execution time of typically a few minutes for a 10~h LHC fill on a standard desktop PC, without notable loss in precision.

To model the transverse and longitudinal motion, the particle coordinates are updated deterministically through a one-turn map based on the machine tune, chromaticity and longitudinal parameters such as RF voltage and momentum compaction~\cite{bruce10prstabCTE}. Any particles outside of the stable area of the RF bucket or a user-defined transverse aperture cut are removed. CTE also includes linear betatron coupling in a thin-lens approximation. 

In CTE, radiation damping and quantum excitation in each plane $u$ are modelled as a combination of  a random excitation and a deterministic decay, with the latter given by the emittance damping time $T_{\mathrm{rad},u}$. We expand the damping coefficient to first order in $T_\mathrm{rev}/T_{\mathrm{rad},u}$ with $T_\mathrm{rev}$ being the revolution time, as described in detail in Ref~\cite{bruce10prstabCTE,siemann85}. It should be noted that in the LHC, the blowup from IBS is much stronger than the quantum excitation. 

\subsection{Beam-beam collisions}

To model the collisions between two bunches, we calculate a collision probability $P_1$ for each macro particle, and a sampled random number determines whether the particle is removed. 
$P_1$ can be calculated either through a numerical solution of the full overlap integral of the single particle trajectory with the opposing bunch distribution, or through a faster approximate method, where the opposing transverse bunch distribution is fitted to a Gaussian, although the actual longitudinal distribution is still accounted for. This allows some of the integrals to be evaluated analytically, resulting in a better computational performance. The detailed derivations and resulting equations are given in Ref.~\cite{bruce10prstabCTE}.

The collision schedule, determined by the bunch filling pattern, can be modelled either through a fast and simplified approach, where the full beam is represented by one macro bunch and the collision probability at each IP is scaled by the number of real bunches colliding, or through a more detailed, but slower approach where one macro bunch is tracked per beam and per beam-beam equivalence class, as defined in~\cite{jowett99}. For the CTE studies in this article, we rely on the first approach with one macro bunch per beam and the approximated $P_1$, since it provides speed and simplicity in the setup and, as will be shown in Sec.~\ref{sec:2018}, the results are nevertheless in very good agreement with LHC data. 

The effect of emittance blowup from core depletion in the collisions, first discovered in Refs.~\cite{bruce10prstabCTE,core-depletion} and further developed in Ref.~\cite{tomas20_em_growth_collisions}, is automatically accounted for. As in the real machine, the collision probability in CTE is higher for macro particles in the core, meaning that more particles are removed in the core than in the tails, and hence the effective emittance grows. 

\subsection{Intra-beam scattering}

The effect of IBS is modelled by giving each macro particle a random kick per turn. The distribution of the kicks is inferred from the instantaneous growth times in the transverse and longitudinal planes, which are re-calculated on every turn. The growth times are first calculated approximating the bunch distributions as Gaussian, which allows the use of the well-established calculations    detailed below. However, in order to account for non-Gaussian longitudinal profiles, the longitudinal kicks are modulated by the local particle density following the method in Refs.~\cite{blaskiewicz-cool07,blaskiewicz08}. 
This allows the   evolution of highly non-Gaussian longitudinal bunch profiles, as  in  RHIC~\cite{bruce10prstabCTE}. 
Optionally the user can specify a mixing factor between the horizontal and vertical growth rates; as this factor is small although not well known for the LHC, we have generally set it to zero.

Several IBS models are available to choose from, 
which we denote 
\begin{itemize}
    \item \emph{Piwinski smooth}~\cite{piwinski74}, 
\item \emph{Piwinski lattice}~\cite{piwinski74}, similar to  \emph{Piwinski smooth}  but using the full lattice and optics instead of a smooth approximation,
\item \emph{modified Piwinski}~\cite{bane02}, 
\item \emph{Bane}~\cite{Bane:2002sr}, a high-energy approximation of the well-known 
\item \emph{Bjorken-Mtingwa} model~\cite{bjorken83}, 
\item \emph{Nagaitsev}~\cite{nagaitsev05}, a re-formulation of \emph{Bjorken-Mtingwa}, with some assumptions that allow faster numerical calculations.
\item the user can also provide an external file with pre-calculated IBS growth rates from any external model, sampled on a grid of bunch dimension values, from which CTE makes an online interpolation. 
\end{itemize}

\begin{table}[tb]
\begin{center}
\caption{\label{tab:ibs} Calculated IBS rise times $T_\mathrm{IBS}$ for the emittances $\epsilon$ (with rise times for the bunch dimensions being a factor~2 larger) for the Pb beam parameters in Table~\ref{tab:lhc} assumed for future LHC runs. Results are shown for the longitudinal (\Tpibs) and horizontal (\Txibs) planes, and obtained with  the models included in CTE and MBS, as well as with standalone calculations using the full \emph{Bjorken-Mtingwa} model or with MAD-X. The rise times are defined by $d\epsilon/dt=\epsilon/T_\mathrm{IBS}$.
}
\begin{tabular}{lrr}
\hline
                                & \Tpibs (h)  & \Txibs (h) \\ \hline 
CTE \emph{Piwinski smooth}    &  2.86 &   4.77        \\ 
CTE \emph{Piwinski lattice}   &  3.15 &   6.88        \\   
CTE \emph{Modified Piwinski}  &  3.19 &   6.40        \\ 
CTE \emph{Bane} &  3.17 &   6.36        \\ 
CTE \emph{Nagaitsev}    &  3.22 &   6.44        \\ 
MBS \emph{CIMP}         &  2.79  &  5.57       \\ 
\emph{Bjorken-Mtingwa }  &  3.21 &   6.41        \\
\emph{MAD-X}            & 3.06  & 6.13          \\ \hline
\end{tabular}
\end{center}
\end{table}

An in-depth comparison of the IBS models is beyond the scope of this paper, but we show as an example in Table~\ref{tab:ibs} the  growth times obtained for the future Pb beam parameters at the start of a fill in Table~\ref{tab:lhc}. The CTE calculations are compared with off-line calculations using both the MAD-X program~\cite{madx}, which uses a development of the Conte-Martini model~\cite{martini85}, and a standalone implementation of  \emph{Bjorken-Mtingwa}. 
It can be seen that the latter gives very similar results to both \emph{Bane} and \emph{Nagaitsev}, as well as  \emph{modified Piwinski}. 
The growth time in the longitudinal plane is about a factor~2 shorter than in the horizontal plane.  
Most models with flat, uncoupled beam optics give a very long negative vertical growth time which is typically unimportant for the dynamics and not given. 
For the sake of comparison, the energy spread has been forced to the value in Table~\ref{tab:lhc} in all computations in Table~\ref{tab:ibs}, while the energy spread calculated internally during CTE simulations is typically slightly shorter, since CTE exactly matches the longitudinal phase space based on the input bunch length and the longitudinal Hamiltonian. 
For all the studies shown in this paper, we have used the \emph{Nagaitsev} model. 

\section{Multi-Bunch Simulation (MBS)}
\label{sec:mbs}

The second simulation code, MBS, is based on a very large set of coupled ODEs, which model the evolution of the bunch parameters of  every  single bunch in both beams. It can therefore model the real collision schedule according to the applied filling scheme. On the other hand, it   relies on the assumption of Gaussian bunch distributions, as opposed to CTE where the tracked particles can assume any distribution. 

For bunch $j$, the ODEs for the bunch intensity $N_j$ and  RMS beam dimensions $\sigma_{u,j}$ in plane $u$ are implemented as~\cite{marc-thesis}:
\begin{equation}
\label{eq:MBS-Nj}
\dot{N}_j(t)=- \sum_{i\in \mathrm{IPs} }\left( \sigma\Lu_{ij} + \frac{N_j (t)}{T_\mathrm{IP,i}}\right) - \frac{N_j (t)}{ \Tnc} \, ,
\end{equation}
\begin{equation}
\label{eq:MBS-sigu}
 \dot{\sigma}_{u,j}(t)=\left(\frac{1}{2 T_{u,\mathrm{IBS}}} - \frac{1}{2 T_{u,\mathrm{rad}}} \right)\sigma_{u,j}(t) \, .
\end{equation}
Equation~(\ref{eq:MBS-sigu}) holds for the planes $u=x,y,z$, giving in total four ODEs per bunch, which translates into more than 5000~ODEs for the fully filled LHC machine. 
As can be seen in Eqs.~(\ref{eq:MBS-Nj})--(\ref{eq:MBS-sigu}), MBS includes the effects of luminosity burn-off, IBS through the rise time $\Tuibs$, radiation damping through the damping time $\Tradu$, and generic extra losses through a lifetime $\Tnc$ for all bunches and $T_\mathrm{IP,i}$ for IP-dependent losses. As we define $\Tradu$ and $\Tuibs$ to refer to changes in the  emittances, a factor~2 is included in Eq.~(\ref{eq:MBS-sigu}) to get the corresponding changes for the bunch dimensions. 
Note that \Lum\ is calculated per IP and bunch and is a function of the parameters of the collision partner for bunch $j$ at IP $i$, denoted by the \textasciitilde\ symbol and subscript $ij$ as
\begin{equation}
 \Lu_{ij}=\Lu_{ij} (\sigma_{x,j},\sigma_{y,j},\sigma_{z,j},N_j,\tilde{\sigma}_{x,ij},\tilde{\sigma}_{y,ij},\tilde{\sigma}_{z,ij} \tilde{N}_{ij}).
\end{equation}
This mechanism effectively couples the evolution of all bunches via Eq.~(\ref{eq:MBS-Nj}). 
The determination of the collision partners relies on the input of the full filling pattern. 

MBS calculates $\Tradu$ with the usual formalism including the radiation integrals~\cite{helm73,handbook98}, while $\Tuibs$ is calculated with the so-called \emph{completely integrated modified Piwinski model} (CIMP)~\cite{kubo05,mtingwa08}. It accounts for the full lattice and can be used for very fast numerical evaluations. Earlier comparisons have shown a very good agreement with  \emph{Bjorken-Mtingwa}~\cite{kubo05}. As seen in Table~\ref{tab:ibs}, it gives for the LHC a similar longitudinal growth as the \emph{Nagaitsev} model in CTE, while the horizontal growth time is about 10\% shorter, meaning a slightly stronger IBS effect. The user can apply ad-hoc correction factors to manually alter the IBS strength and a parameter to mix the horizontal and vertical growth rates. 
This has not been done in this article. 

In MBS, Eqs.~(\ref{eq:MBS-Nj})--(\ref{eq:MBS-sigu}) are integrated numerically using the explicit Euler algorithm as discussed in Ref.~\cite{marc-thesis}. The time steps $\Delta t$ can be specified and in this work we use $\Delta t=\qty{3}{min}$.

Luminosity levelling can be activated as in CTE, in which a scaling factor $S_i=\text{min}(1,\Lu_{ti}/\Lu_{\mathrm{pot},i})$ is applied to $\Lu_{ij}$ in Eq.~(\ref{eq:MBS-Nj}) to keep it at a constant, user-defined target $ \Lu_{ti}$ at IP $i$. Here $\Lu_{\mathrm{pot},i}$ is the maximum potential luminosity at IP $i$ calculated over all bunches.

Additional, generic losses can be accounted for through the non-collisional lifetime $\Tnc$. Bunches colliding at IP $i$ can be assigned an extra IP-specific lifetime $T_\mathrm{IP,i}$, which has been shown to increase the agreement with early data~\cite{marc-thesis}. 

Several physical effects are neglected in both MBS and CTE. Firstly, beam-gas collisions could, apart from the risk of increased experimental backgrounds~\cite{bruce13_NIM_backgrounds,bruce19_PRAB_beam-halo_backgrounds_ATLAS}, lead to both emittance growth and intensity decay~\cite{handbook98}. However, with the very good vacuum levels achieved in the LHC, these processes are weak compared to the strong burn-off in the collisions and the emittance growth from IBS. 
However beam-gas interactions are likely responsible for a substantial part of the observed non-collisional losses discussed in Sec.~\ref{sec:2018} and  can easily be included in both CTE and MBS as a generic non-collisional lifetime. 

Other influences on the  beam include noise in feedback systems, RF cavities or magnet power supplies, and  dynamic aperture losses due to magnet non-linearities and beam-beam effects. 
These effects depend on imperfections that are not well-known in quantitative detail. As the simulation result for Pb operation is dominated by burn-off and very close to the measurements, as will be shown in Sec.~\ref{sec:2018}, we conclude that these effect are minor in  LHC heavy-ion operation.   
Other studies suggest that they may be important for protons~\cite{papadopoulou19_evian}. 
Nevertheless, losses due to these additional effects can be subsumed in the generic non-collisional lifetime in both CTE and MBS, and any effect on the emittance can be included in the generic emittance risetime in CTE.

\section{Analysis of the 2018 Pb-Pb run}
\label{sec:2018}

\subsection{Simulation setup}

\begin{table*}
  \centering
    \begin{tabular}{lrrr}     \hline        
                & \qty{6.37\,Z}{TeV} & \qty{7\,Z}{TeV}  & \qty{7\,Z}{TeV} \\ 
                & Pb-Pb        & Pb-Pb                  & p-Pb \\ \hline 
Hadronic inelastic (b)   & 7.7  & 7.8 & 2.13 \\ 
Bound-free pair production---BFPP (b)  & 278  & 281 & 0.044 \\ 
Electromagnetic dissociation---EMD  (b)  & 223  & 226 & 0.035 \\ 
Total(b)  & 509  & 515 & 2.21 \\ \hline                
    \end{tabular}%
   \caption{Burn-off cross sections for various interactions between colliding Pb-Pb beams at \qty{6.37\,Z}{TeV}, as in the 2018 LHC operation, for future Pb-Pb operation at \qty{7\,Z}{TeV}, and for future \qty{7\,Z}{TeV}7   p-Pb collisions. 
   The \qty{7\,Z}{TeV} Pb-Pb values are obtained from Refs.~\cite{lhcdesignV1,meier01,pshen01,baltz99,dEnterria18_hadronic_cross_sections} and the \qty{6.37\,Z}{TeV}  ones are estimated using a scaling by the fixed-target frame $\gamma$ of $\log(2\gamma^2-1)$. 
   In accordance with Refs.~\cite{meier01,dEnterria18_hadronic_cross_sections,smirnov14_prab}, such a scaling is very close to the complete calculation. 
   The p-Pb values are taken from~\cite{marc-thesis}. 
   The EMD cross sections include all decay channels. }
  \label{tab:cross_sections}%
\end{table*}

To compare the simulations with data, 30~out of the 46 physics fills in the 2018 LHC Pb-Pb run~\cite{jowett19_ipac} at a beam energy of \qty{6.37\,Z}{TeV} were simulated in detail with CTE and MBS. The remaining 16 fills were de-selected due to either missing logged data, very short fill lengths, or non-standard operational procedures interfering with the evolution. The simulations cover the so-called stable-beam period in each fill, which starts at the point in the operational cycle where the beams are brought into collision and left evolving and it continues until the fill is terminated by a beam dump. 

The starting conditions for the simulations were taken from the 2018 operational parameters in Table~\ref{tab:lhc}, except for the parameters that varied between fills, which were instead extracted from logged data. The starting intensities, emittances, and bunch lengths varied due to fluctuations in the injector performance as well as small variations between injection and stable beams in the LHC. For CTE, using the approach with a single macro bunch per beam, the average bunch parameters were used, while for MBS, individual values were extracted for each bunch. Furthermore, the filling scheme and hence the number of bunches colliding at each IP was changed several times in order to first have a gradual ramp-up and then deploy the new 75~ns scheme in the second half of the run. The luminosity levelling target was also gradually increased from one fill to the next, in order to carefully probe the limits from losses due to BFPP. 

From the starting point of the colliding beams, the simulated beam parameters and luminosity evolved independently and no further input from measurements was used. Since non-colliding bunches in the machine showed on average 100~h lifetime, we applied the same non-collisional lifetime in the simulations but did not include any IP-dependent lifetimes in MBS or any additional emittance blowup. A total burn-off cross section $\sigmatot=\qty{509}{b}$ for particle removal was assumed, consisting of the contributions shown in Table~\ref{tab:cross_sections}. As can be seen, it is dominated by the electromagnetic processes (BFPP and EMD), and the hadronic part is a minor contribution. 

Because of the very large \sigmatot,  
the LHC Pb ion operation is in a strong burn-off regime, where the total number of injected Pb ions, $N_{1,2}$ in either beam,    determines the maximum possible integrated luminosity, 
\begin{equation}
\label{eq:maxLum} 
 \sum_{\mathrm{expts}} \int_{0}^{\Tf} \Lu(t)\, dt \le \frac{\min(N_1,N_2)}{\sigmatot}
\end{equation}
where equality is approached in the limit of vanishing non-collisional losses and exhaustion of the
lesser beam as the fill length $\Tf\to\infty$.  
In typical LHC heavy-ion fills the ratio of the two sides of this inequality,  the \emph{luminous efficiency},  exceeds 50--60\%.

During the first half of the 2018 run, there was an error in the local coupling correction around IP2~\cite{persson19_evian}, which   significantly reduced the ALICE luminosity but was later corrected. For the affected fills we used an effective \bstar-value of 0.9~m in the simulations. This value, which   reproduces  the measured luminosity in the simulated fills, is  based on the luminosity scans in~\cite{wenninger_LMC18}. It is the \bstar\ that would give the same emittance in the ALICE scans as the value inferred from the scans at ATLAS and CMS.

\begin{figure*}[!tbhp]
    \centering
    \includegraphics[width=0.49\textwidth]{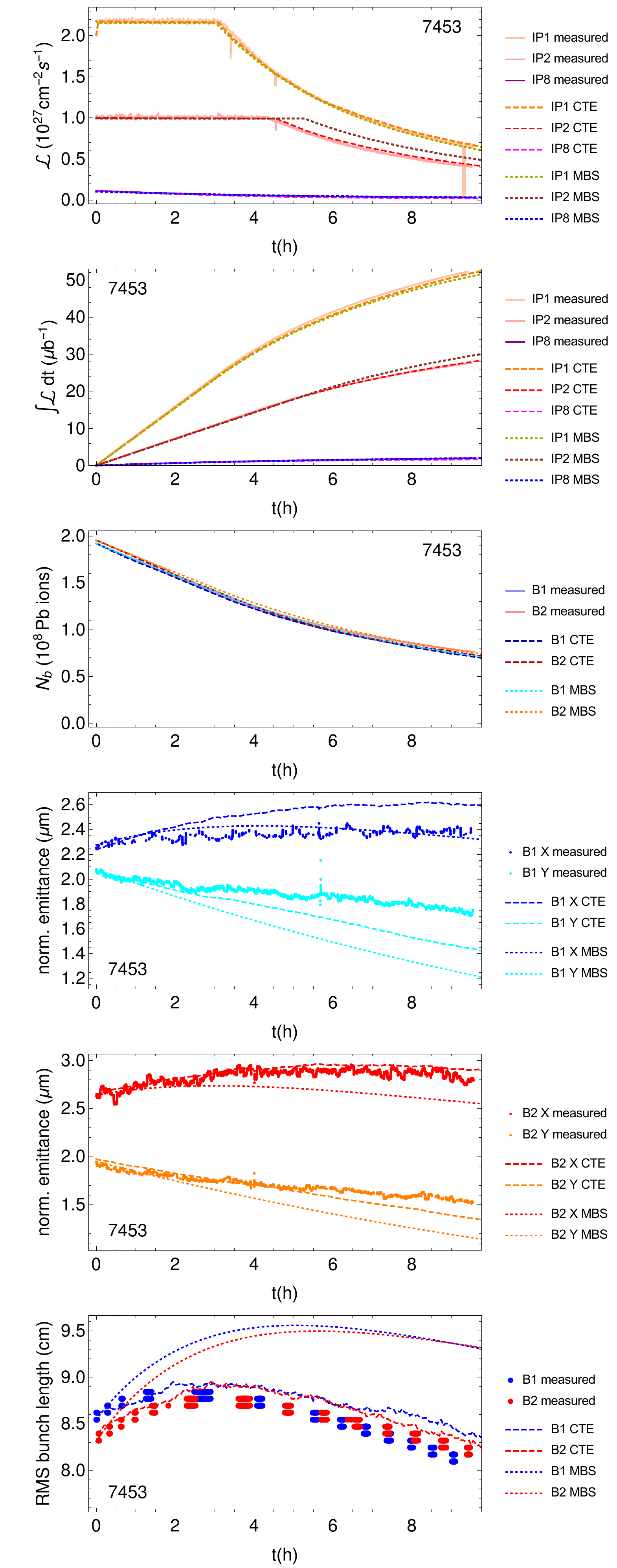}
    \includegraphics[width=0.49\textwidth]{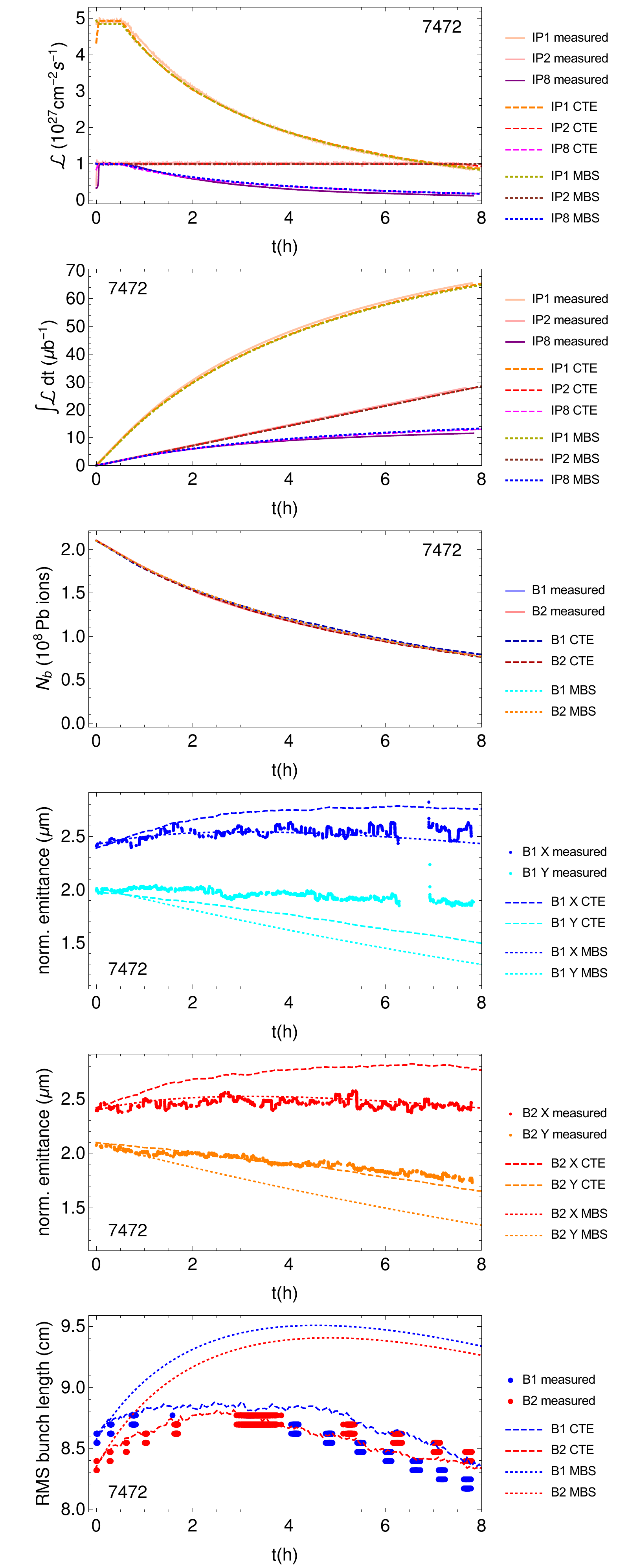}
    \caption{The measured evolution (solid lines) of key observables (instantaneous luminosity, integrated luminosity, average bunch intensity $N_b$, and average transverse emittances and bunch lengths) during two typical Pb-Pb fills (7453--left, 7472--right) from the 2018 LHC run, compared to simulation results from CTE  (dashed lines) and MBS (dotted lines). The IP5 luminosity is not shown as it is almost identical to the one in IP1. }
    \label{fig:comp_2018}
\end{figure*}

\begin{figure*}[!tbhp]
    \centering
    \includegraphics[width=0.49\textwidth]{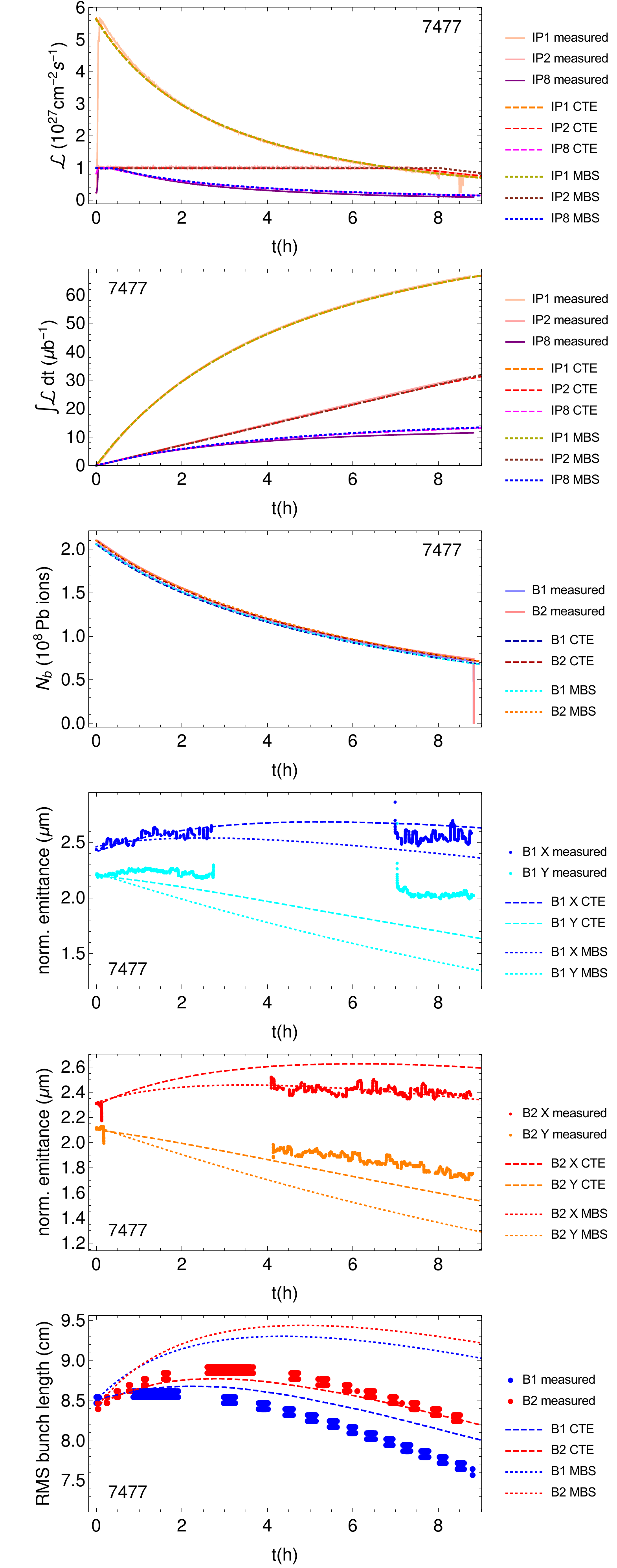}
    \includegraphics[width=0.49\textwidth]{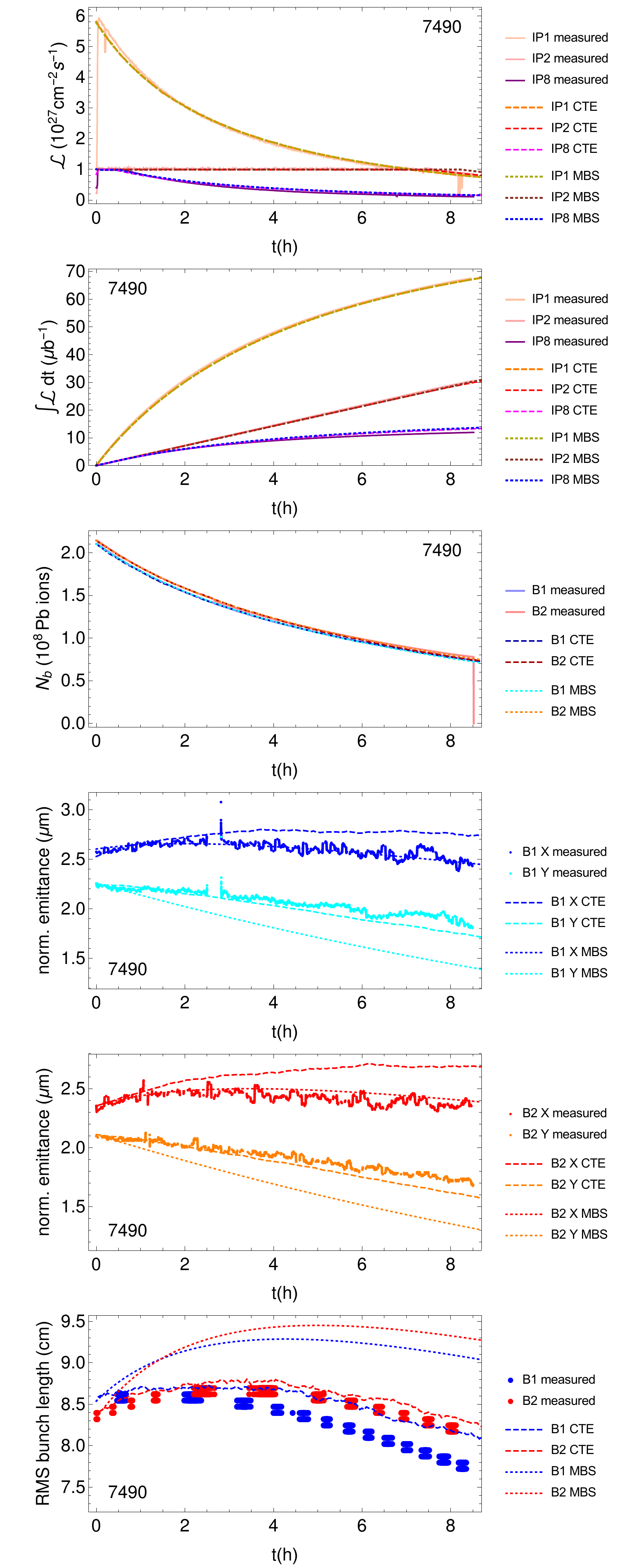}
    \caption{The measured evolution (solid lines) of key observables (instantaneous luminosity, integrated luminosity, average bunch intensity $N_b$, and average transverse emittances and bunch lengths) during two typical Pb-Pb fills (7477--left, 7490--right) from the 2018 LHC run, compared to simulation results from CTE  (dashed lines) and MBS (dotted lines). The IP5 luminosity is not shown as it is almost identical to the one in IP1. }
    \label{fig:comp_2018_2}
\end{figure*}

\subsection{Results}

Results for a few typical fills with different starting conditions are presented in Figs.~\ref{fig:comp_2018}--\ref{fig:comp_2018_2}. For each fill, the simulated and measured online evolution of the instantaneous and integrated luminosity can be seen, as well as the beam intensity, and an excellent agreement is found in these quantities. It should be noted though that the end of the luminosity levelling at IP2 occurs slightly later in MBS than in CTE or the logged data. The other simulated fills are very similar in the level of agreement. 

A fair agreement is found also in the emittance and bunch length evolutions. 
This comes in spite of an uncertainty on the measured emittances from the synchrotron light monitor (BSRT), which was never calibrated in detail for Pb beams in the 2018 run, and which sometimes has missing data as,    e.g., in fill~7477 in Fig.~\ref{fig:comp_2018_2}. Nevertheless, using the measured emittance values as initial conditions at the start of each fill results in a very good agreement in the key observables in Fig.~\ref{fig:comp_2018},  strongly suggesting that the real error on the emittance measurement is small. 

\begin{figure}[!tbhp]
    \centering
    \includegraphics[width=0.7\textwidth]{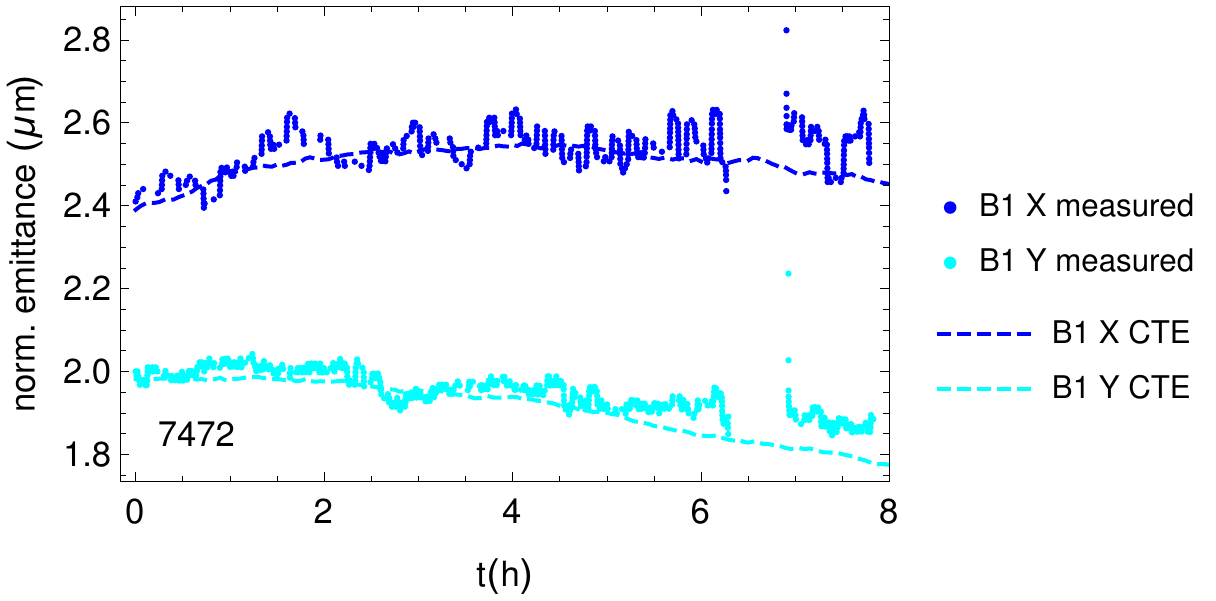}
    \caption{Example of simulated emittances from CTE for fill~7472, shown in Fig.~\ref{fig:comp_2018}, but with using an empirically fitted mixing value between the IBS growth rates in the horizontal and vertical planes. We used $T_{x,\mathrm{IBS,eff}}$\xspace=0.65\Txibs + 0.35 \Tyibs with the analogous mixing in the vertical plane.  }
    \label{fig:ibs_fitted}
\end{figure}

The horizontal emittance generally grows initially, while the vertical emittance shrinks, since radiation damping dominates in the vertical plane. 
The simulated horizontal emittance from CTE typically grows more than the measured one, while the simulated vertical emittance shrinks more. This is likely due to the assumptions of zero IBS mixing between the planes in the simulation. An IBS mixing value can be empirically fitted to generally yield a good agreement with the measured emittances as shown in Fig.~\ref{fig:ibs_fitted} and without significantly affecting the simulated luminosity. 
However, as the real value in the machine is unknown we have refrained from doing so in the other results. 

\begin{table*}
  \centering
  
    \begin{tabular}{lrrrr}     \hline        
            & CTE 2018  & MBS 2018  & CTE future    & MBS future \\ \hline 
\Tpibs (h)  & 3.4       & 3.7       & 2.8           & 2.8 \\         
\Txibs (h)  & 8.8       & 8.4       & 6.2           & 5.6 \\ 
\Tradp (h)  & -8.5      & -8.5      & -6.4          & -6.4 \\ 
\Tradx (h)  & -16.9     & -16.9     & -12.4         & 12.4 \\ 
\Tcore (h)  & 28        & ---       & 23            & --- \\ 
\Teff  (h)  & 11.1      & 16.7      &  8.1          & 10.2 \\ \hline 
\end{tabular}%
   \caption{Calculated emittance rise times and damping times for IBS, radiation damping, and core depletion, as well as the total effective rise time over all processes $i$, calculated as $1/T_\mathrm{eff}=\sum_i 1/T_i$. The results refer to the start of collisions for the typical 2018 and future configurations in Table~\ref{tab:lhc}. }
  \label{tab:risetimes}%
\end{table*}

In MBS, the emittances stay smaller than in CTE, in spite of a slightly stronger IBS growth (see Table~\ref{tab:ibs}). This is due to the core depletion, as can be understood from Table~\ref{tab:risetimes}, where we compare the strengths of the main effects. For the horizontal plane in a typical 75~ns fill with 733~bunches, IBS is largely counteracted by radiation damping at the start of the fill, with a residual combined effective risetime \Teff$\approx$17--18~h. However, if the  core depletion rise time of about 28~h is also included, the effective total growth time goes down to about 11~h. Since core depletion is included by construction in the CTE collision routine, the effective emittance   grows much more than in MBS which does not include this effect. On the other hand, the bunch length grows significantly more in MBS than in CTE, with CTE showing a better agreement with the measurements. This is driven mainly by stronger longitudinal IBS due to the smaller transverse emittance. 

\begin{figure}[!tbh]
    \centering
    \includegraphics[width=0.49\textwidth]{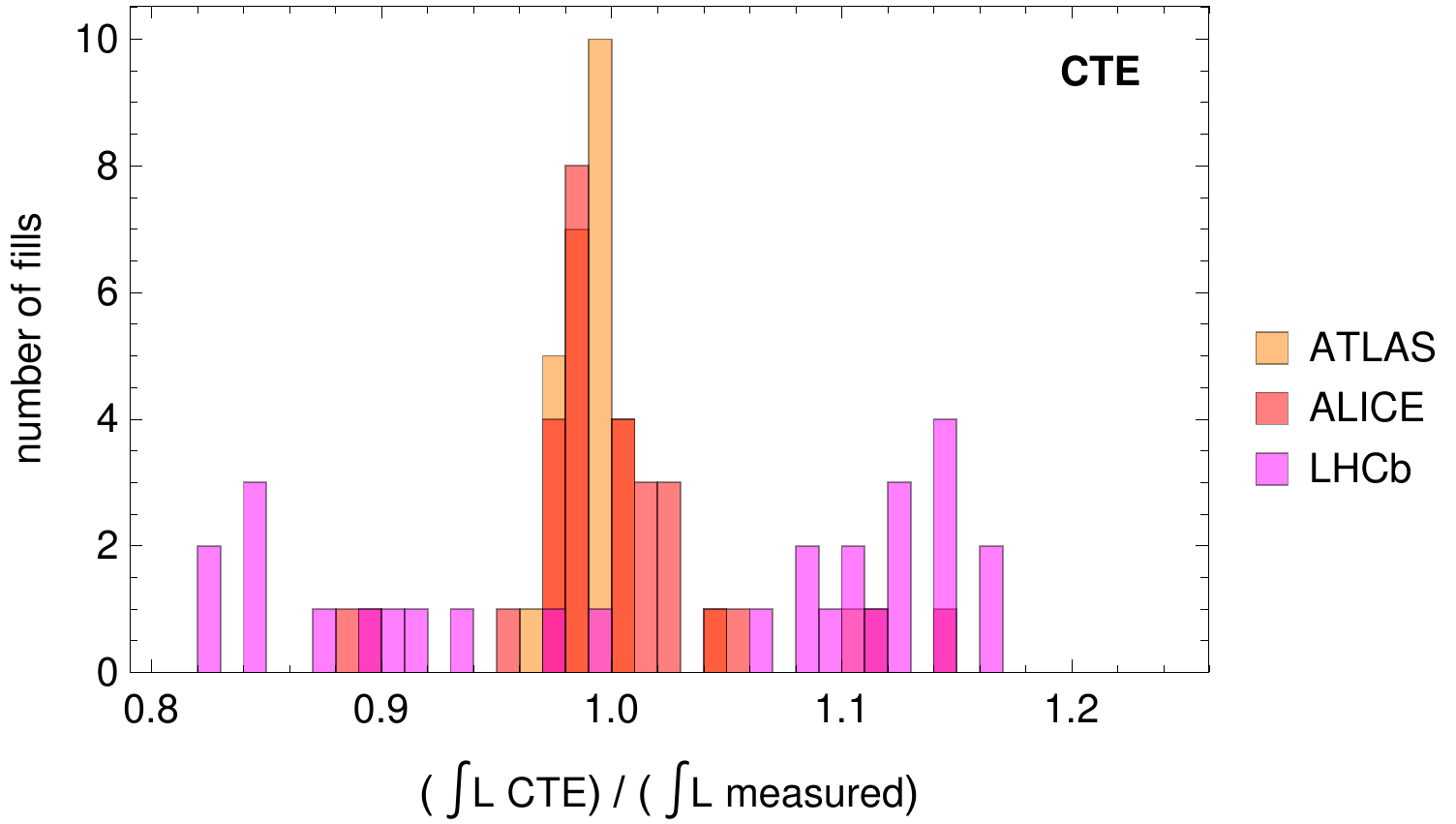}
    \includegraphics[width=0.49\textwidth]{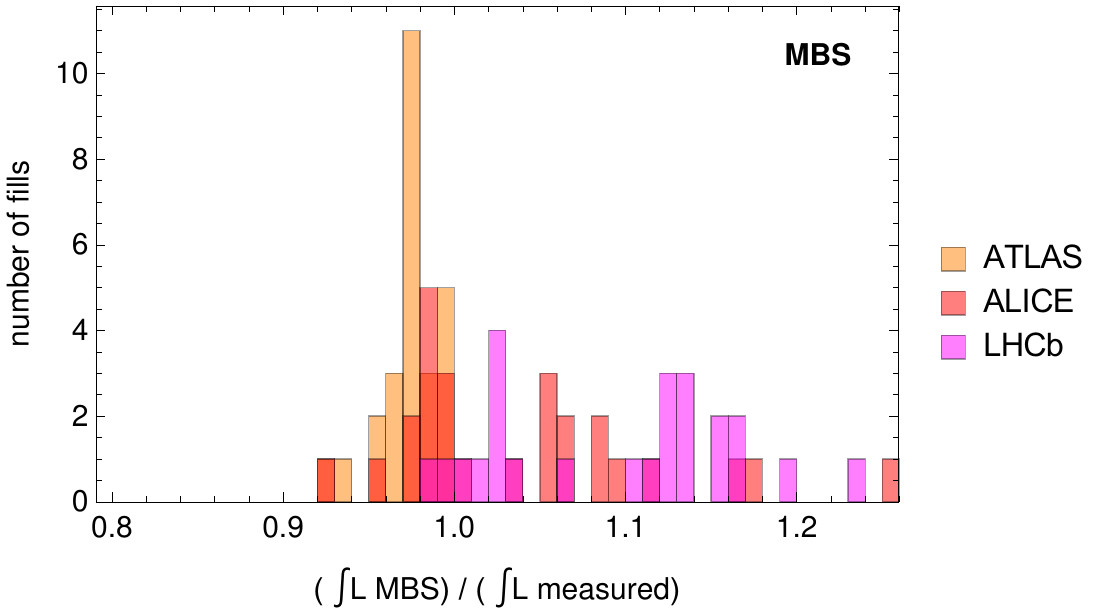}
    \caption{Distribution of the ratio of simulated (with CTE---left, MBS---right) to measured integrated luminosity for ATLAS, ALICE and LHCb per fill in the 2018 Pb-Pb run.  
    } 
    \label{fig:hist_2018}
\end{figure}

In practical terms, the most important quantity for the simulation benchmark is the integrated luminosity at the end of each fill. The ratios of \intlum between simulations and measurements in the 30 simulated fills are shown in Fig.~\ref{fig:hist_2018}, and the mean and standard deviations per experiment   in Table~\ref{tab:lum_mean_sigma}. 
For the fills analysed and for both codes, the average ratio is very close to 1 in both IP1 and IP2, with a small standard deviation of a few percent, which we consider an excellent agreement.

At IP8, a the spread is much larger, with a standard deviation of 0.12. The detailed reason for this is not fully clear, although it is likely related to very large uncertainties in the LHCb luminosity calibration~\cite{FAlessio-private}. MBS consistently overestimates \intlum at IP8. In CTE, it is underestimated for a couple of fills in the early part of the run where very few bunches collided at IP8, while the luminosity in later fills with more colliding bunches is overestimated. This is possibly due to larger differences between individual bunches for the schemes with few collisions, that can only be seen with the MBS approach. The two codes are roughly consistent for all fills with a significant number of collisions at IP8.

In spite of the small discrepancies at IP8, we consider the overall agreement very good. The fact that both codes largely agree, in spite of being based on different underlying principles, strengthens our confidence in the results and shows that they can be used for reliable predictions of the luminosity in future machine configurations. 

\begin{table}
  \centering

    \begin{tabular}{lrrr}     \hline        
            & IP1  & IP2  & IP8  \\ \hline 
CTE, mean  & 1.00  & 1.01 & 1.02           \\ 
CTE, RMS   & 0.03  & 0.04 & 0.12           \\ 
MBS, mean  & 0.98  & 1.04 & 1.09           \\ 
MBS, RMS   & 0.01  & 0.06 & 0.07           \\ \hline                
\end{tabular}%
   \caption{Mean and RMS value of the ratio between simulated and measured integrated luminosity over the 30~analyzed fills in the LHC 2018 Pb-Pb run. Results are shown for both the CTE and MBS codes and for the different experiments. The results for IP5 are not shown as they are very similar to IP1.}
  \label{tab:lum_mean_sigma}%
\end{table}

\section{Projected performance in future runs}
\label{sec:hl-lhc}

In this section we estimate the heavy-ion performance in future LHC runs, based on simulations of typical fills using CTE and MBS, which are extrapolated to a typical one-month run. We discuss Pb-Pb operation in Sec.~\ref{sec:PbPbHL_perf} and p-Pb in Sec.~\ref{sec:pPbHL_perf}.

\subsection{Pb-Pb performance in the HL-LHC baseline configuration}  
\label{sec:PbPbHL_perf}

We assume the future running scenario detailed in Ref.~\cite{bruce20_HL_ion_report}, with the key parameters shown in Table~\ref{tab:lhc}. Some of the most notable differences to the 2018 configuration are the increased number of bunches and collisions, the higher beam energy and the higher luminosity levelling target at IP2 of \lum{6.4}. We assume the same levelling at IP1 and IP5 for simplicity, although the experiments there could potentially accept a higher luminosity. At IP8, we assume levelling at \lum{1} to avoid quenches from BFPP. We use \sigmatot=515~b at \qty{7\,Z}{TeV}, with details given in Table~\ref{tab:cross_sections}. As can be seen in Table~\ref{tab:risetimes}, the processes influencing the emittances are stronger in the future configuration, with an effective initial risetime of about 8~h if core depletion is included. 

\begin{figure*}[!tbh]
    \centering
    \includegraphics[trim={0.0cm 0 0 0},clip,width=\textwidth]{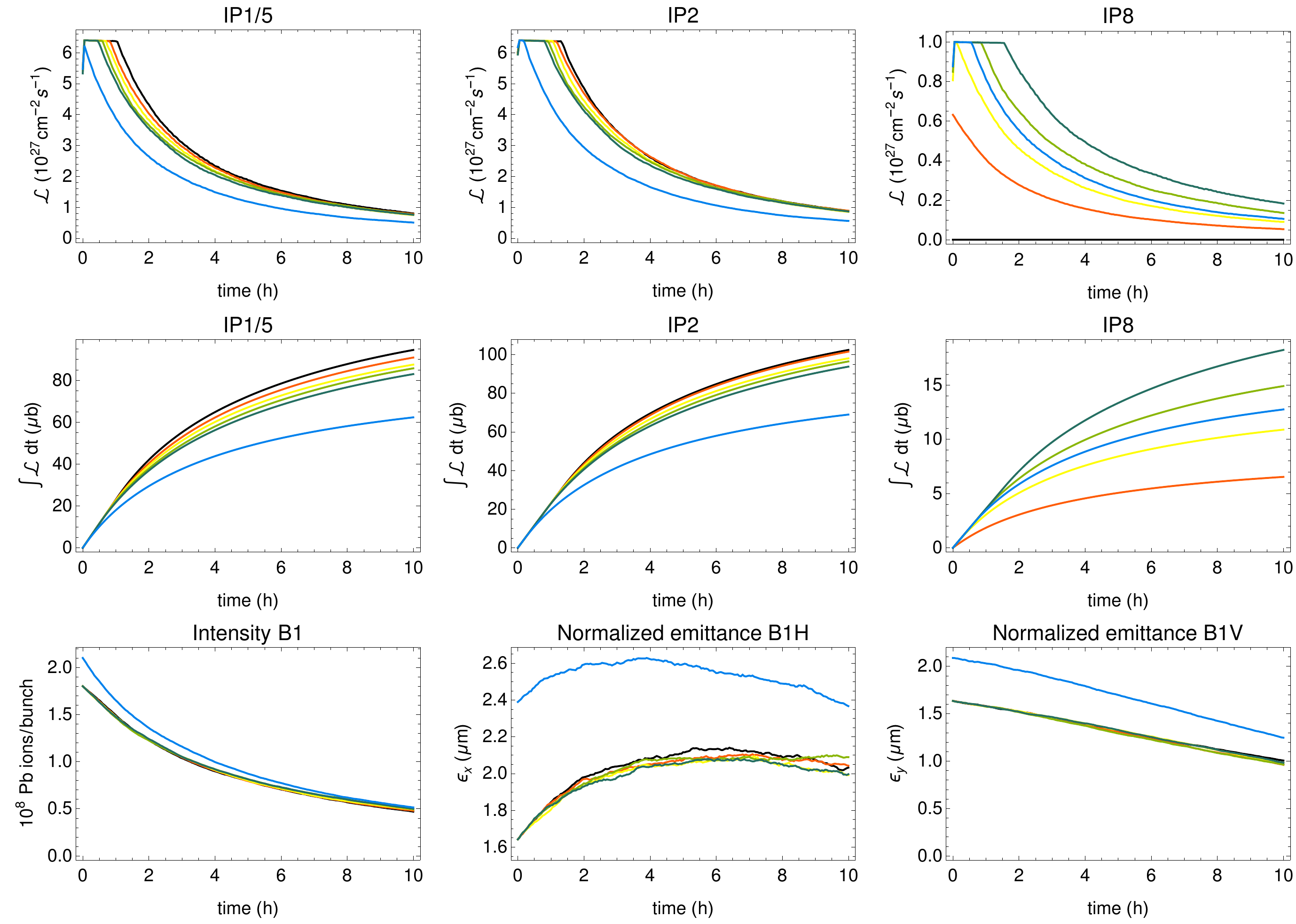}
    \includegraphics[trim={0 0 0 6.3cm},clip,width=0.6\textwidth]{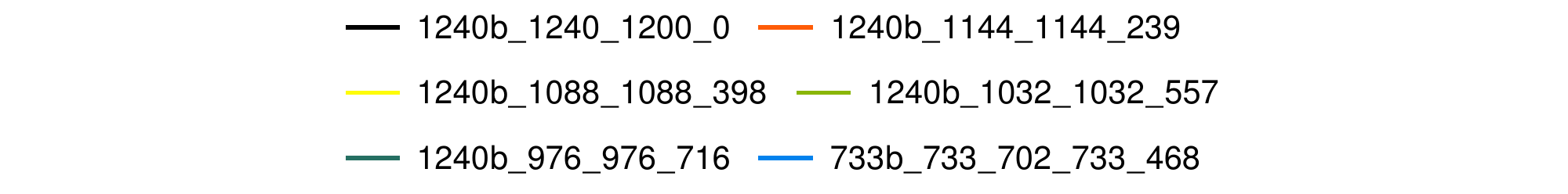} 
    \caption{ The simulated HL-LHC Pb-Pb performance from CTE in terms of instantaneous luminosity (top) and integrated luminosity (middle) during a typical fill for the considered filling schemes from Ref.~\cite{bruce20_HL_ion_report}, shown together with the evolution of the beam intensity and normalized emittances (bottom). Only B1 is shown, but B2 is fully symmetric.  }
    \label{fig:HL-sim-PbPb}
\end{figure*}

We show the simulated luminosity, intensity, and emittance evolutions for a typical future Pb-Pb fill in Fig.~\ref{fig:HL-sim-PbPb}, with various proposed 50~ns filling schemes from Ref.~\cite{bruce20_HL_ion_report}. The filling scheme names are defined with the customary LHC convention indicating the total number of bunches (b) followed by the number of collisions at ATLAS/CMS (IP1/5), ALICE (IP2), and LHCb (IP8). All new schemes, motivated by a recent request by LHCb for significant Pb-Pb data taking~\cite{YR_WG5_2018}, use 1240~bunches but with different distributions of collisions between  IP1, IP2, and IP5, on the one hand and  IP8 on the other. 
Note that IP8 is slightly displaced from the symmetry point of the LHC ring and therefore requires displaced 50~ns bunch trains to have collisions at the expense of fewer collisions at the other IPs. 

In addition to the 50~ns schemes with 1240~bunches, we simulate also the 75~ns filling scheme \fs{733}{733}{702}{468} from 2018, which is a backup in case of production problems of the 50~ns beams. For this scheme, we assume the 2018 starting conditions in Table~\ref{tab:lhc},  except that we use a \qty{7\,Z}{TeV} energy. 

As can be seen in Fig.~\ref{fig:HL-sim-PbPb}, the intensities and emittances are similar between different 50~ns filling schemes, while different initial parameters are assumed for the 75~ns scheme. The levelling time at IP1 and IP5 is up to about 1~h. IP2 has a slightly longer levelling time because of its smaller crossing angle.
At IP8, levelling is only needed for the three filling schemes with most collisions there. 
The IP8 luminosity is   lower than at the other IPs because of its lower levelling target, fewer  colliding bunches, the larger \bstar, and  larger  crossing angle. 
The integrated luminosity improves significantly at IP8 for filling schemes that provide it with more collisions, at the price of a moderate reduction for the other experiments. As expected, the 75~ns scheme performs significantly worse than the 50~ns schemes. 

As usual, the optimal fill length (time spent in collision), $T_\mathrm{f}$,  can be calculated to maximise the average luminosity $\Lavg$ 
\begin{equation}
\label{eq:avgLum}
 \Lavg (\Tf)\,=\, \frac{\int_{0}^{\Tf} \Lu(t)\, dt} {\Tf+\Tta },
\end{equation}
where $\Lu(t)$ is the instantaneous luminosity given by the simulation, and $\Tta$ is the turnaround time, i.e., the time between the dump and the start of the collisions in the next fill. 
In reality, the turn-around time is the sum of a minimum irreducible value and a random value whose distribution reflects the general operational efficiency but cannot generally be predicted before the previous fill is dumped. 
However for the sake of simplicity and  consistency with the corresponding treatment for protons~\cite{hl-lhc-tech-design,Metral:2301292}, we assume a typical value of $\Tta=200$~min, based on the detailed time estimates in Ref.~\cite{jowett17_cham}.

\begin{figure}[tb]
    \centering
    \includegraphics[width=0.7\textwidth]{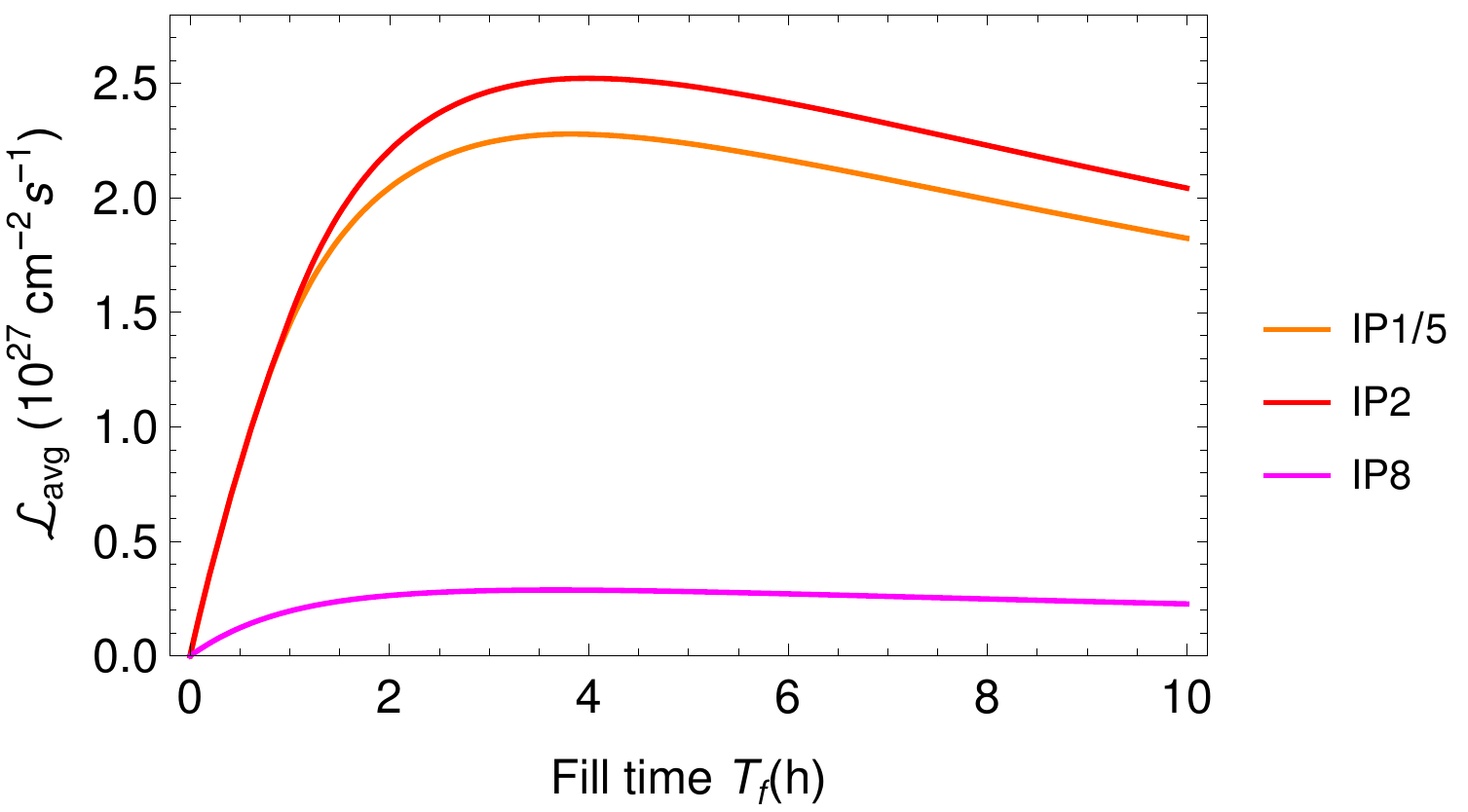}
    \caption{ The time-averaged luminosity as a function of the turnaround time, calculated with Eq.~(\ref{eq:avgLum}) and the CTE simulations in Fig.~\ref{fig:HL-sim-PbPb} for the \fs{1240}{1088}{1088}{398} filling scheme.   }
    \label{fig:avgLum_t}
\end{figure}

As an example, $\Lavg(\Tf)$ is shown in Fig.~\ref{fig:avgLum_t} for the \fs{1240}{1088}{1088}{398} scheme. 
It turns out that the optimum fill time, $\Topt$,   does not differ much between experiments. 
For the   50~ns schemes, it is around 3.8--3.9~h at IP1,  4.0--4.1~h at IP2, and 3.6--4.5~h at IP8. 
The spread in $\Topt$ is largest at IP8, but the curve of $\Lavg$ is also very flat (see Fig.~\ref{fig:avgLum_t}). 
Therefore, for Pb-Pb,  we can adopt the value of  $\Topt$ calculated for IP2 for each filling scheme.

The integrated luminosity $\Ltot$ in one Pb-Pb run is then estimated as 
\begin{equation}
 \Ltot \, = \, \Lavg (\Topt) \times T_\mathrm{run} \times \eta,
 \label{eq:Ltot}
\end{equation}
where $\Trun$ is the total time allocated to the physics run. 
Typically one month per year is allocated for heavy-ion operation.  Allowing  the first week   for commissioning,  we assume $\Trun=24$~days. 

The factor $\eta$ in Eq.~(\ref{eq:avgLum}) is the \emph{operational efficiency}, which should account for downtime and unavailability of the machine, premature fill aborts, occasional longer $\Tta$ and, most importantly, the build-up of performance to the ideal during the few weeks of the run. 
Conventionally, and conservatively, we take $\eta=0.5$, as   for HL-LHC proton operation~\cite{Metral:2301292}. 
In the 2018 heavy-ion run a higher $\eta$ was achieved, when the machine availability was exceptionally 85\% after the initial commissioning~\cite{jowett19_evian}. 
Since $\eta$ takes account of the build-up of luminosity that occurs during these short runs, it is less than the machine availability.  
Note also that   $\eta=0.5$ has been typical during some proton runs.

The calculated $\Ltot$ per one-month run using these parameters and assumptions is shown in Table~\ref{tab:int_lum_PbPb} for the different filling schemes in Ref.~\cite{bruce20_HL_ion_report}. 
Results without brackets are based on CTE, but corresponding MBS simulations have also been performed and these results are shown in square brackets. The results based on the two simulation codes, using fundamentally different models, agree within 5\% except for the filling schemes with very few collisions at IP8. This strengthens our confidence in the results.

For the 50~ns schemes, $\Ltot$ varies in the range 2.2--2.6~\nb at IP1 and IP5, and in the range of 2.4--2.8~\nb at IP2, which has a smaller net crossing angle. 
With the backup 75~ns scheme (last row in Table~\ref{tab:int_lum_PbPb}), the loss in $\Ltot$ at these experiments is 
about 20--30\% per run.  

At LHCb, $\Ltot$ depends strongly on the filling scheme. With the highest  number of collisions considered for 50~ns beams, about 0.5~\nb can be collected per run.
This   is significantly higher than the 0.35~\nb   for the 75~ns backup scheme \fs{733}{733}{702}{468}. 
At the same time, $\Ltot$ at the other experiments is also significantly higher with 50~ns.  
Therefore, we conclude that a 50~ns scheme should always be preferred if available, provided the projected beam quality can be achieved.

\begin{table*}[]
\centering

\begin{tabular}{lccc} \hline 
Filling scheme             & $\Ltot$  IP1/5   & $\Ltot$ IP2             & $\Ltot$ IP8  \\ \hline 
\fs{1240}{1240}{1200}{0}   & 2.5 [2.5]                & 2.7 [2.8]                           & 0 [0]                            \\ 
\fs{1240}{1144}{1144}{239}  & 2.4 [2.4]                & 2.7 [2.7]                           & 0.18 [0.21]                         \\ 
\fs{1240}{1088}{1088}{398} & 2.4 [2.3]                & 2.6 [2.7]                           & 0.30 [0.34]                         \\ 
\fs{1240}{1032}{1032}{557} & 2.3 [2.2]                & 2.5 [2.6]                           & 0.39 [0.44]                         \\ 
\fs{1240}{976}{976}{716} & 2.2 [2.1]                & 2.5 [2.5]                           & 0.46 [0.50]                         \\ 
\fs{733}{733}{702}{468}   & 1.7 [1.7]                & 1.9 [1.9]                           & 0.35 [0.36]                        \\ \hline 
\end{tabular}

\caption{Integrated luminosity (given in \nb) during a one-month Pb-Pb run at each experiment for the considered filling schemes from Ref.~\cite{bruce20_HL_ion_report}, assuming an operational efficiency of $\eta$=0.5~h and  24~days available for physics operation in Eq.~(\ref{eq:avgLum})--(\ref{eq:Ltot}). The first number is calculated using CTE and the number in square brackets is calculated using MBS.   }
\label{tab:int_lum_PbPb}
\end{table*}

\subsection{Sensitivity study and performance enhancements}

\begin{figure}[tb]
    \centering
    \includegraphics[width=0.7\textwidth]{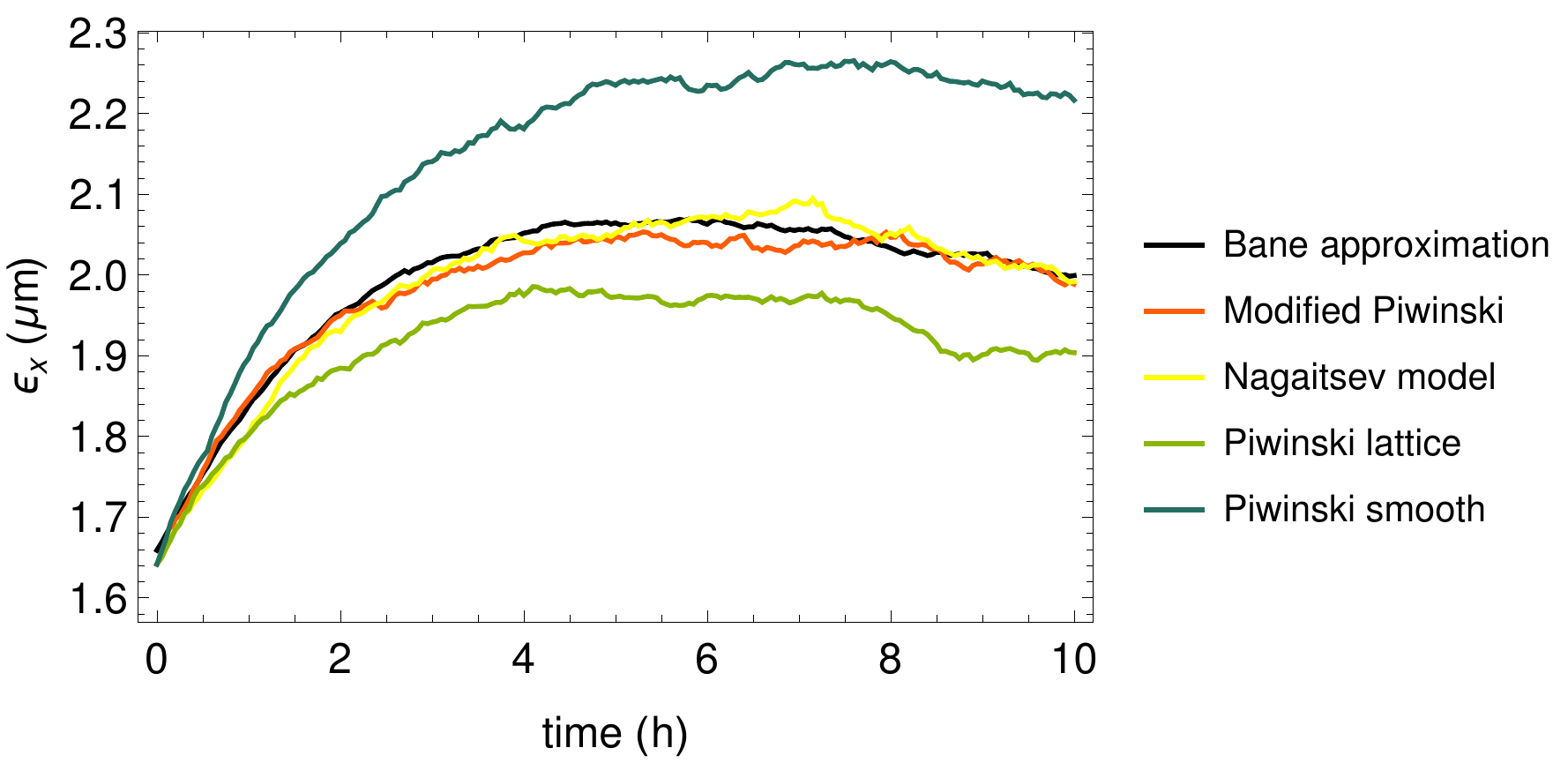}
    \caption{ The evolution of 
    the horizontal emittance during a typical fill as simulated with CTE using different IBS models. The beam parameters for future runs in Table~\ref{tab:lhc} and the filling scheme \fs{1240}{1088}{1088}{398} were assumed.}
    \label{fig:IBS_sensitivity}
\end{figure}

So far we have considered the baseline scenario, but using the simulations and Eq.~(\ref{eq:avgLum})--(\ref{eq:Ltot}) we have also investigated the sensitivity of $\Ltot$ to changes in the baseline parameters and assumptions in the simulations.  One representative filling scheme was selected, \fs{1240}{1088}{1088}{398}, and the inputs were varied in CTE. 
First, we have checked the influence of the IBS model. 
Fig.~\ref{fig:IBS_sensitivity} shows 
the evolution of 
the horizontal emittance for each of the built-in IBS models. 
Minor variations are observed in the emittance.  
The luminosity evolution remains very close to Fig.~\ref{fig:HL-sim-PbPb} in all cases due to the strong burn-off regime and is therefore not shown. 
The calculated $\Ltot$ per one-month run changes by up to 1\% between these runs, where the largest difference is seen for the \emph{Piwinski smooth} model, which is expected to have a poorer accuracy as it does not account for the full lattice. Further studies show that the IBS mixing factor between the transverse planes has an even smaller impact on $\Ltot$.

\begin{figure*}[!tbh]
    \centering
    \includegraphics[width=0.32\textwidth]{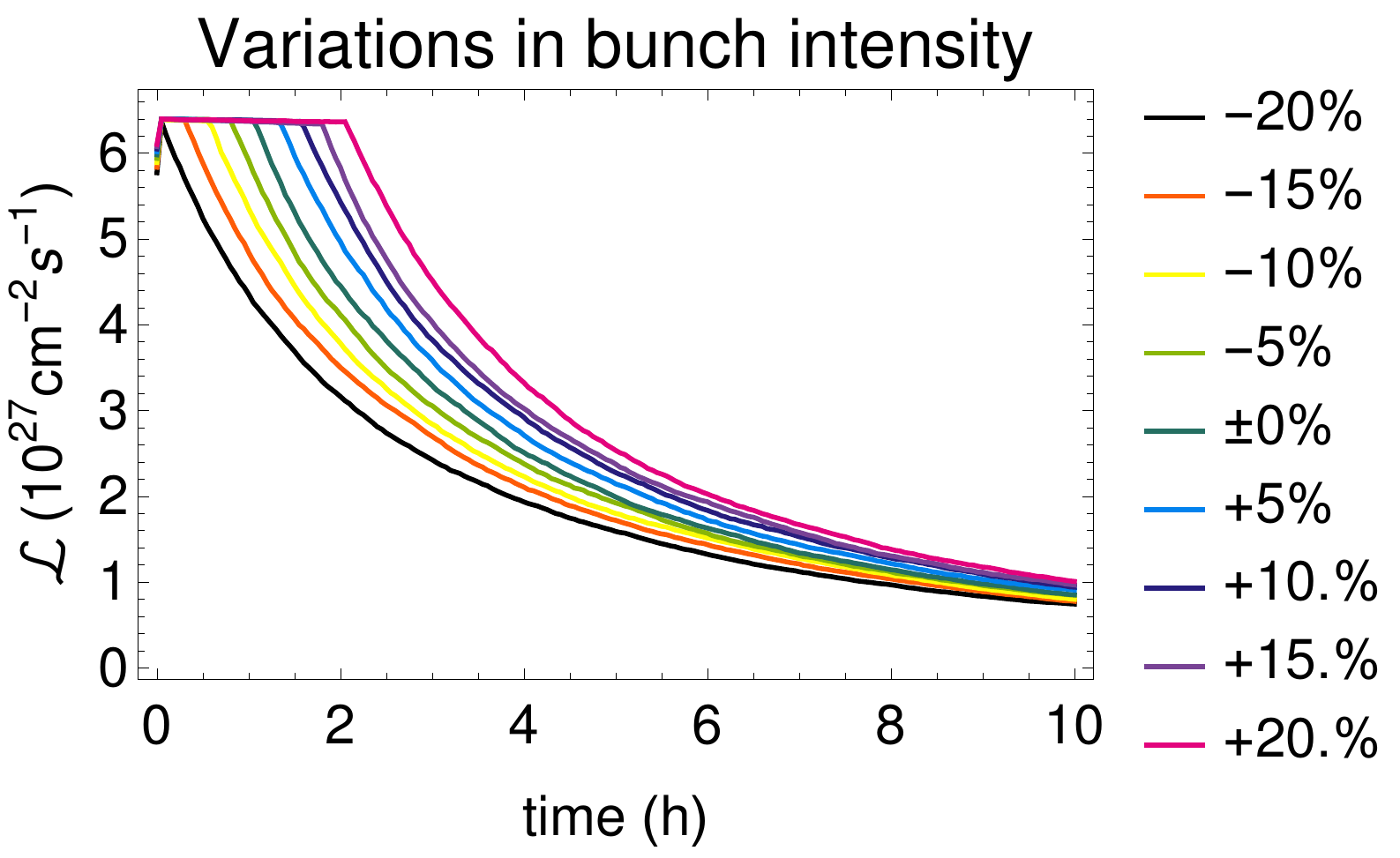}
    \includegraphics[width=0.32\textwidth]{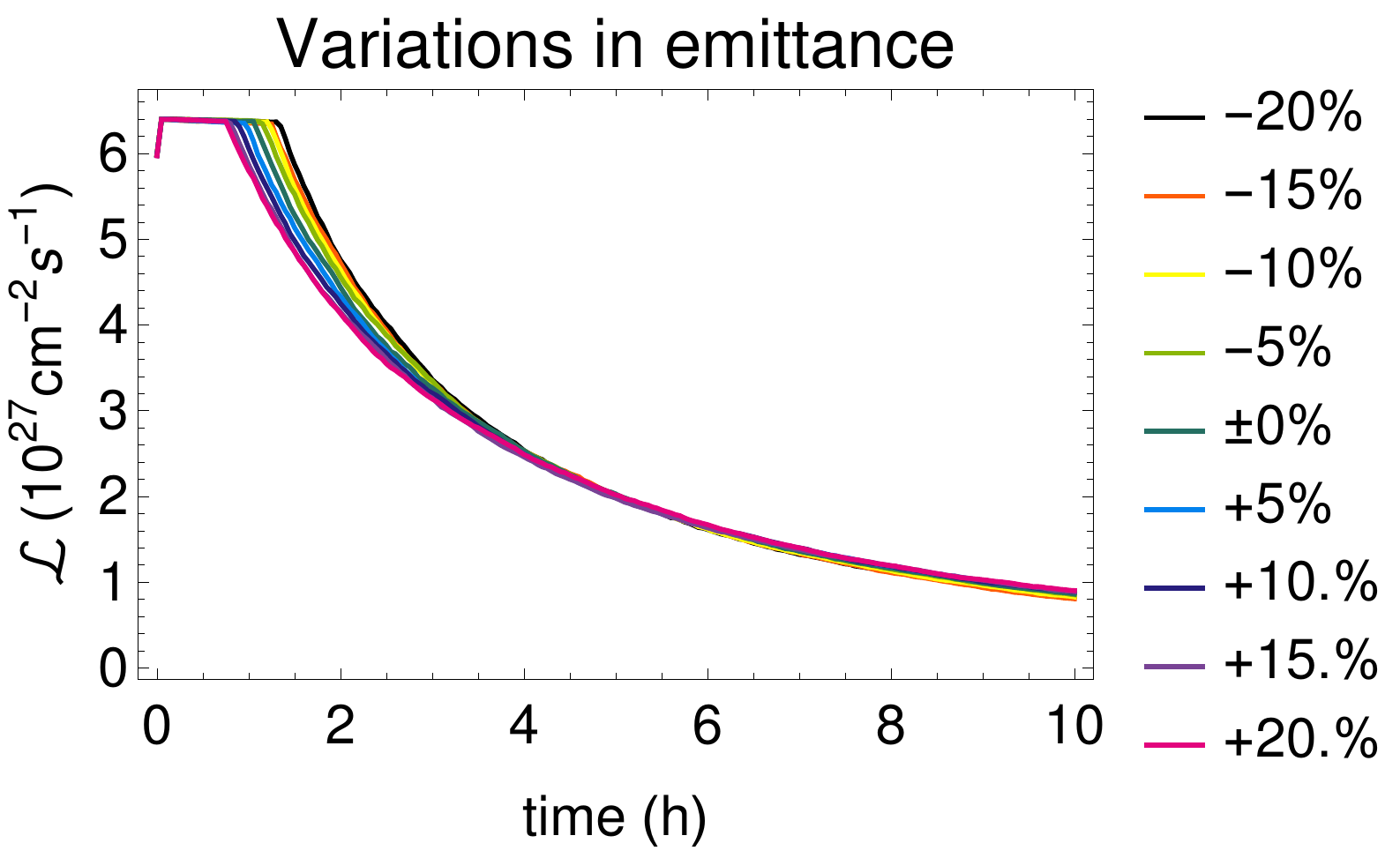}
    \includegraphics[width=0.32\textwidth]{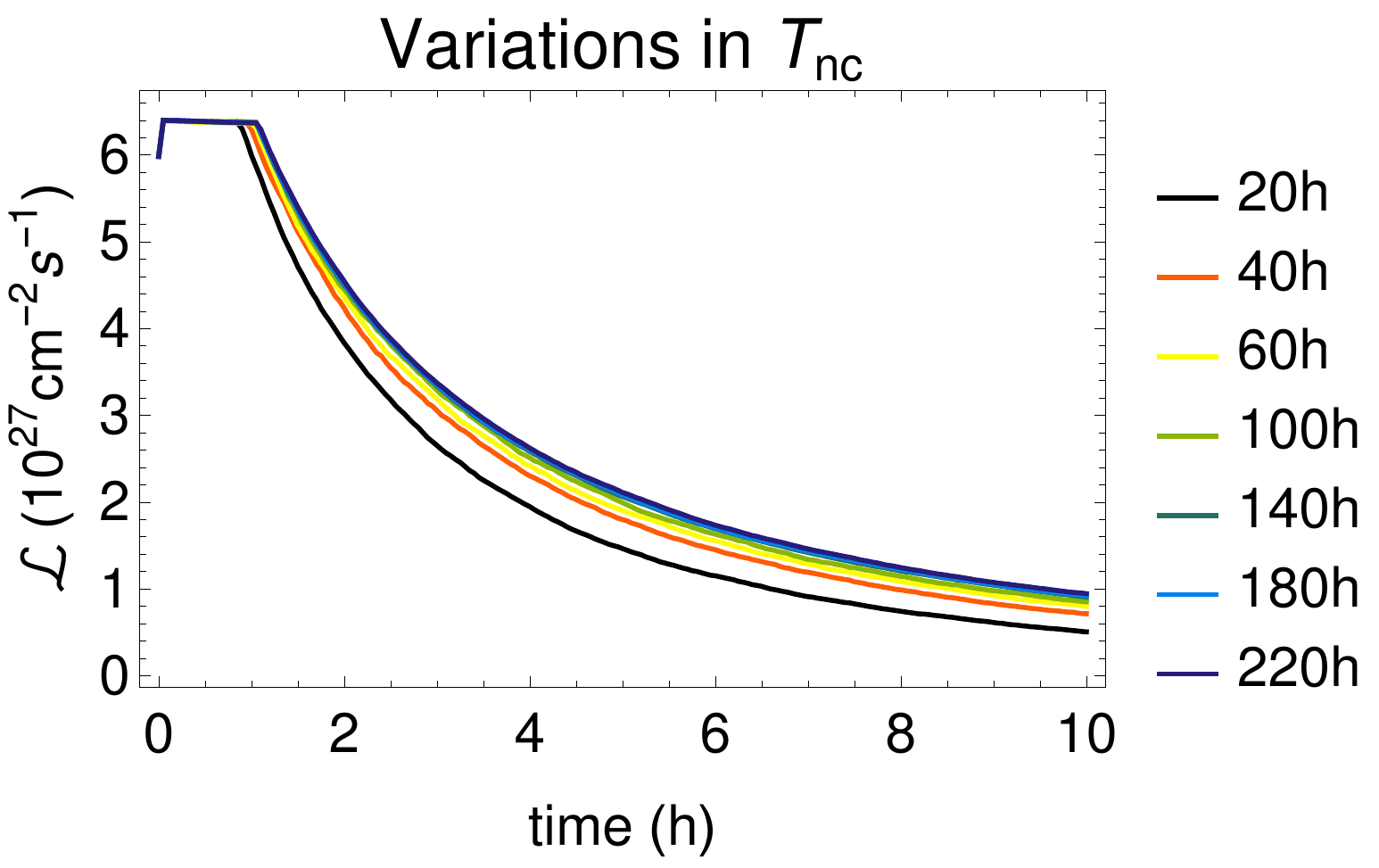}
    \caption{ The evolution of the instantaneous luminosity at IP2 when varying the bunch intensity (left), the transverse emittance (middle), and the non-collisional lifetime $\Tnc$ during a typical fill as simulated with CTE. The beam parameters for future runs in Table~\ref{tab:lhc} and the filling scheme \fs{1240}{1088}{1088}{398} were assumed.}
    \label{fig:lum_sensitivity}
\end{figure*}

\begin{figure*}[!tbh]
    \centering
    \includegraphics[width=0.32\textwidth]{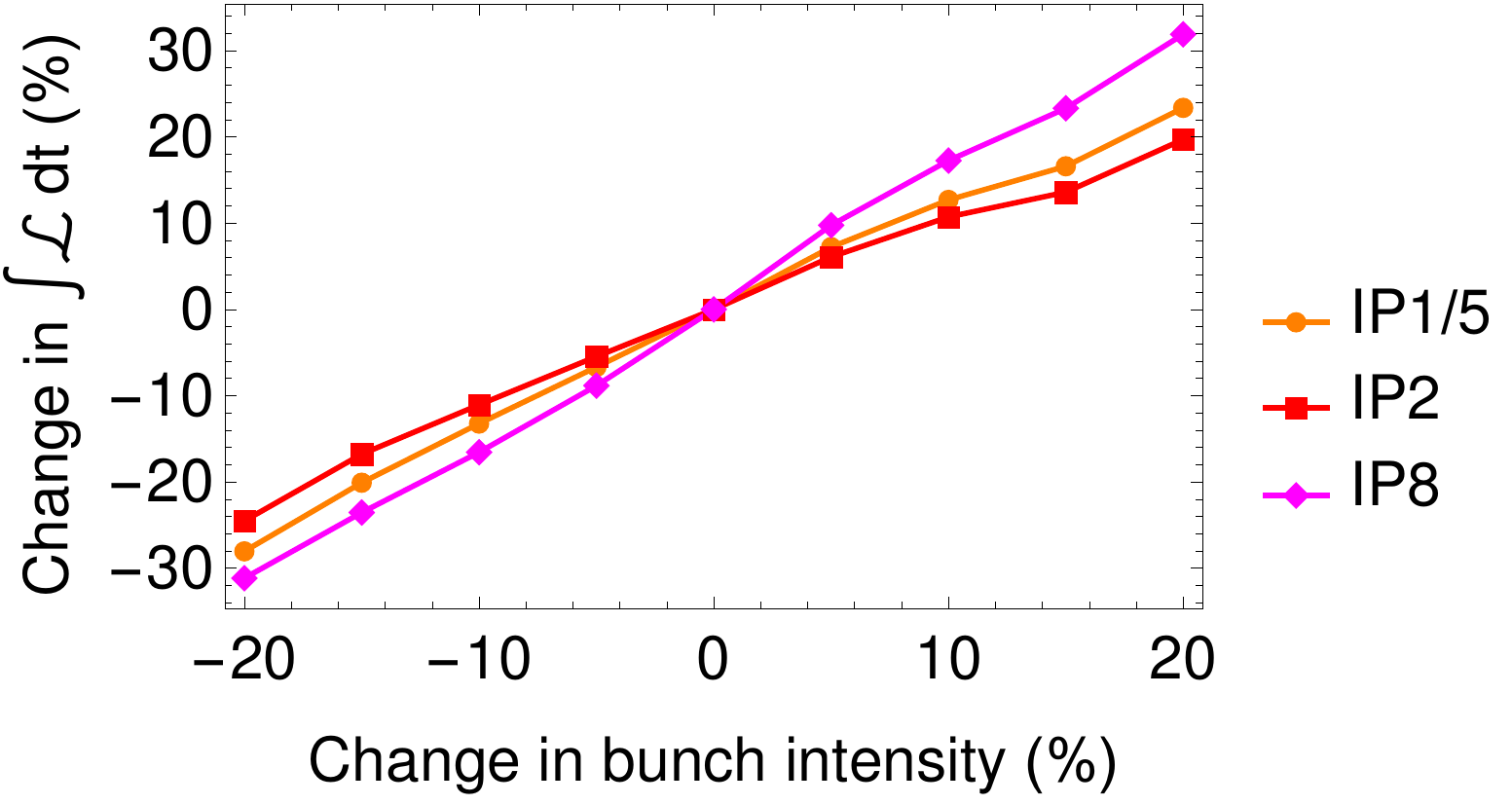}
    \includegraphics[width=0.32\textwidth]{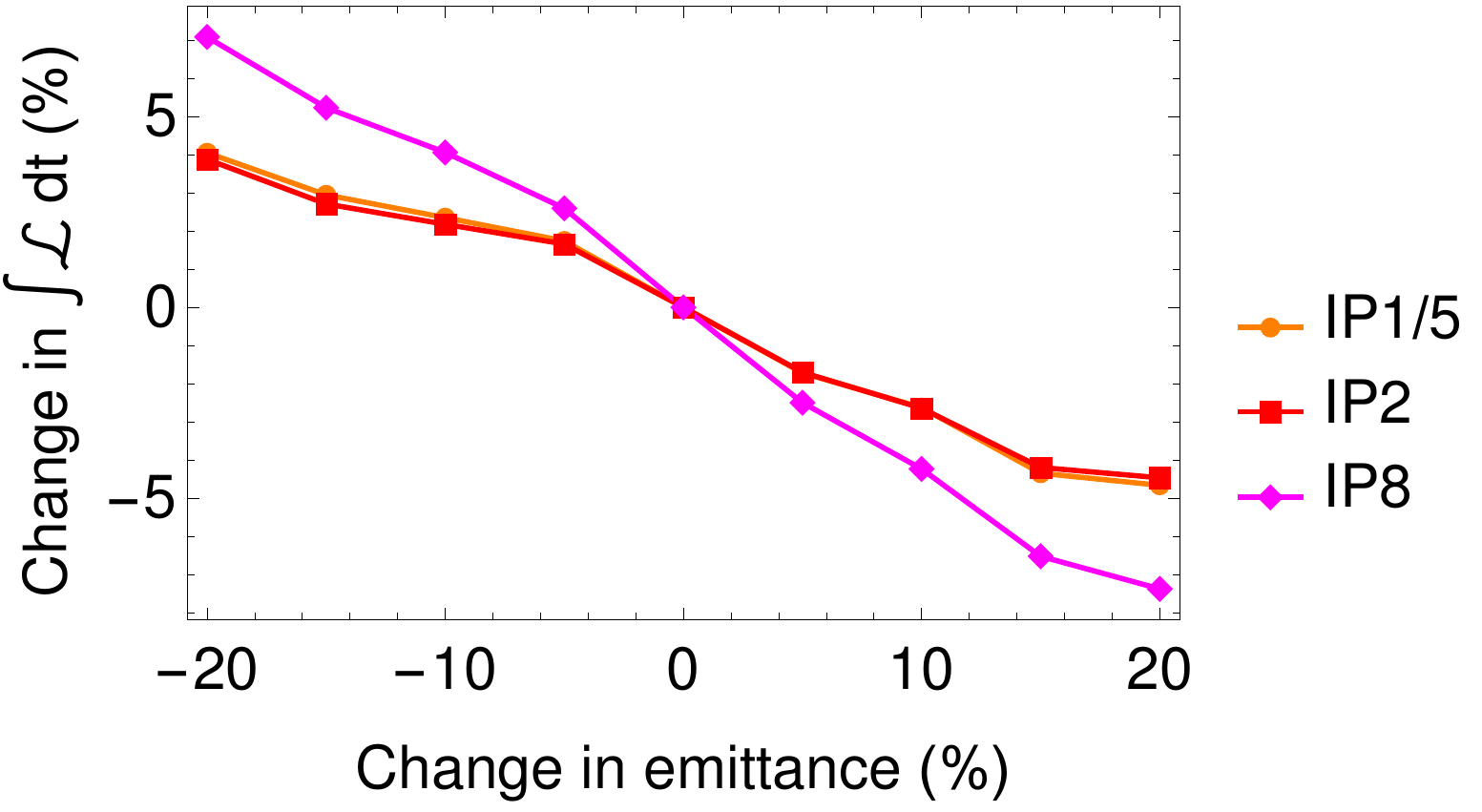}
    \includegraphics[width=0.32\textwidth]{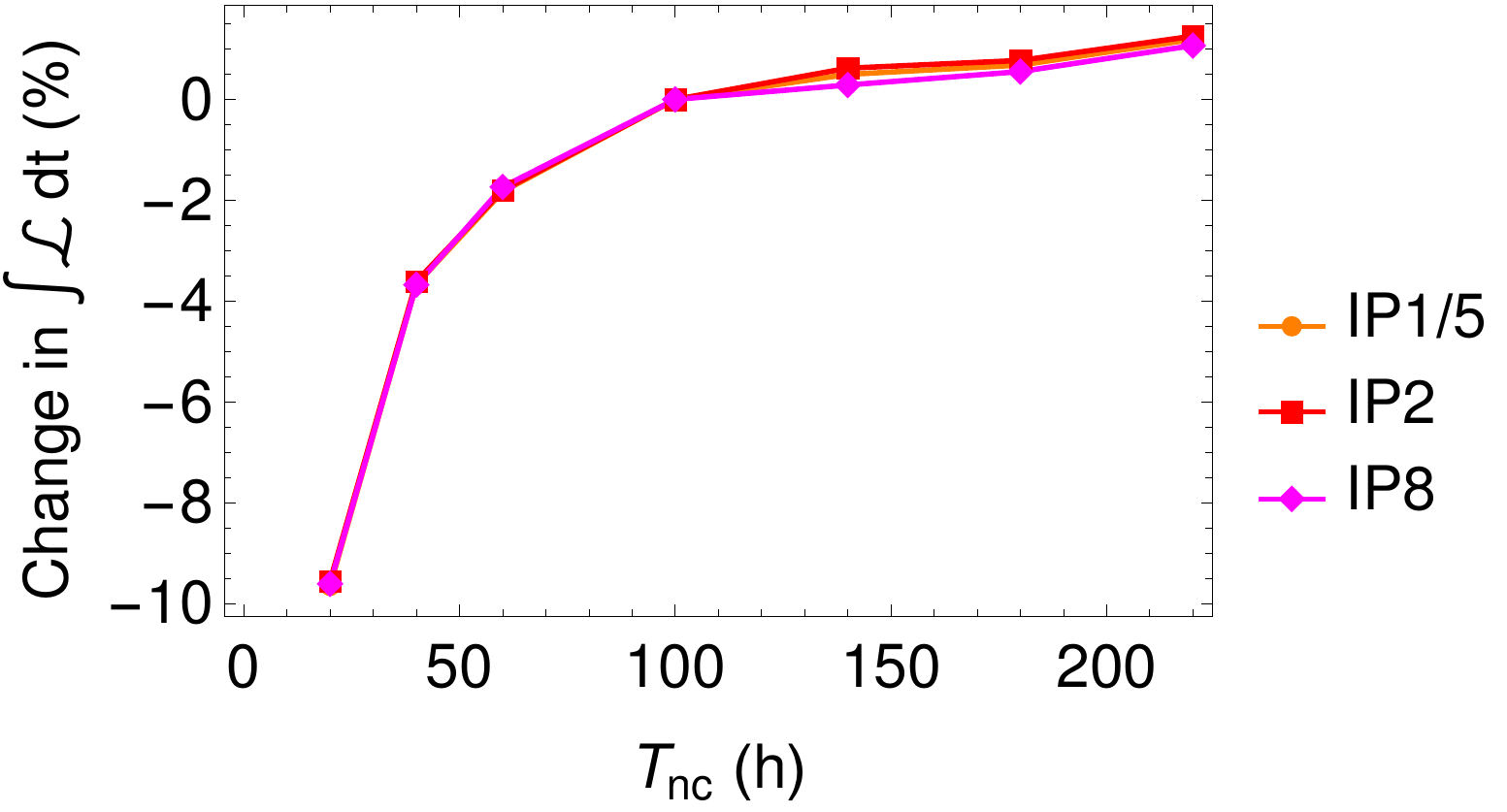}
    \caption{ The change in $\Ltot$ during a full one-month run as a function of the change in bunch intensity (left) and transverse emittance (right) as calculated from CTE simulations. The beam parameters for future runs in Table~\ref{tab:lhc} and the filling scheme \fs{1240}{1088}{1088}{398} were assumed.}
    \label{fig:lum_int_sensitivity}
\end{figure*}

We have also studied the influence of the non-collisional lifetime $\Tnc$, the bunch intensity $N$, and the transverse emittance \exy. So far we used $\Tnc=100$~h from the fit of the non-colliding bunches in the 2018 run. The source of these losses is not fully understood, and a scan was performed over $\Tnc$ between 20~h and 220~h. The bunch intensity and emittance were varied within a $\pm20$\% interval of the design values in Table~\ref{tab:lhc}.

The simulated \Lum\ at IP2 is shown in Fig.~\ref{fig:lum_sensitivity} and the resulting change in $\Ltot$ during a one-month run is shown in Fig.~\ref{fig:lum_int_sensitivity}. Since the beam lifetime from the luminosity burn-off evolves in the range of 5--8~h, the negative influence of a lower $\Tnc$ becomes clearly visible only for very low values, and for $\Tnc>100$~h, the gains of longer lifetimes are very small. The estimated $\Ltot$ changes by less than 2\% for all tested values of $\Tnc$ except $\Tnc=40$~h, giving a 4\% reduction, and $\Tnc=20$~h, which reduces $\Ltot$ by 10\%. We consider that such low values of $\Tnc$ are very unlikely and would only occur in case of serious problems with the machine. 

On the other hand, Figs.~\ref{fig:lum_sensitivity}--\ref{fig:lum_int_sensitivity} show that $N$ strongly affects both \Lum\ and $\Ltot$, although $\Ltot$ is closer to a linear $N$-dependence than the $N^2$ dependence of the instantaneous \Lum. A 10\% change in $N$ results in a 10--14\% change in $\Ltot$. 
As expected, the dependence on \exy\ is weaker, and the variations of $\Ltot$ stay within about 5\% except at IP8, where the effect is slightly larger.

Finally, as can be seen in Eq.~(\ref{eq:Ltot}), $\Ltot$ varies linearly with the operational efficiency $\eta$, which has a rather large uncertainty. 
Gains in $\eta$ could have a very significant impact. 

We have also investigated changes in the machine configuration whose feasibility is yet to be demonstrated.   Nevertheless, these studies can indicate paths to higher luminosity. First, \bstar\ might be squeezed to smaller values. The lower limit, still to be quantified in detail, is given by the protection of the available aperture, as shown for proton operation in Refs.~\cite{bruce15_PRSTAB_betaStar,bruce17_NIM_beta40cm}. 

\begin{figure}[!tbh]
    \centering
    \includegraphics[width=0.7\textwidth]{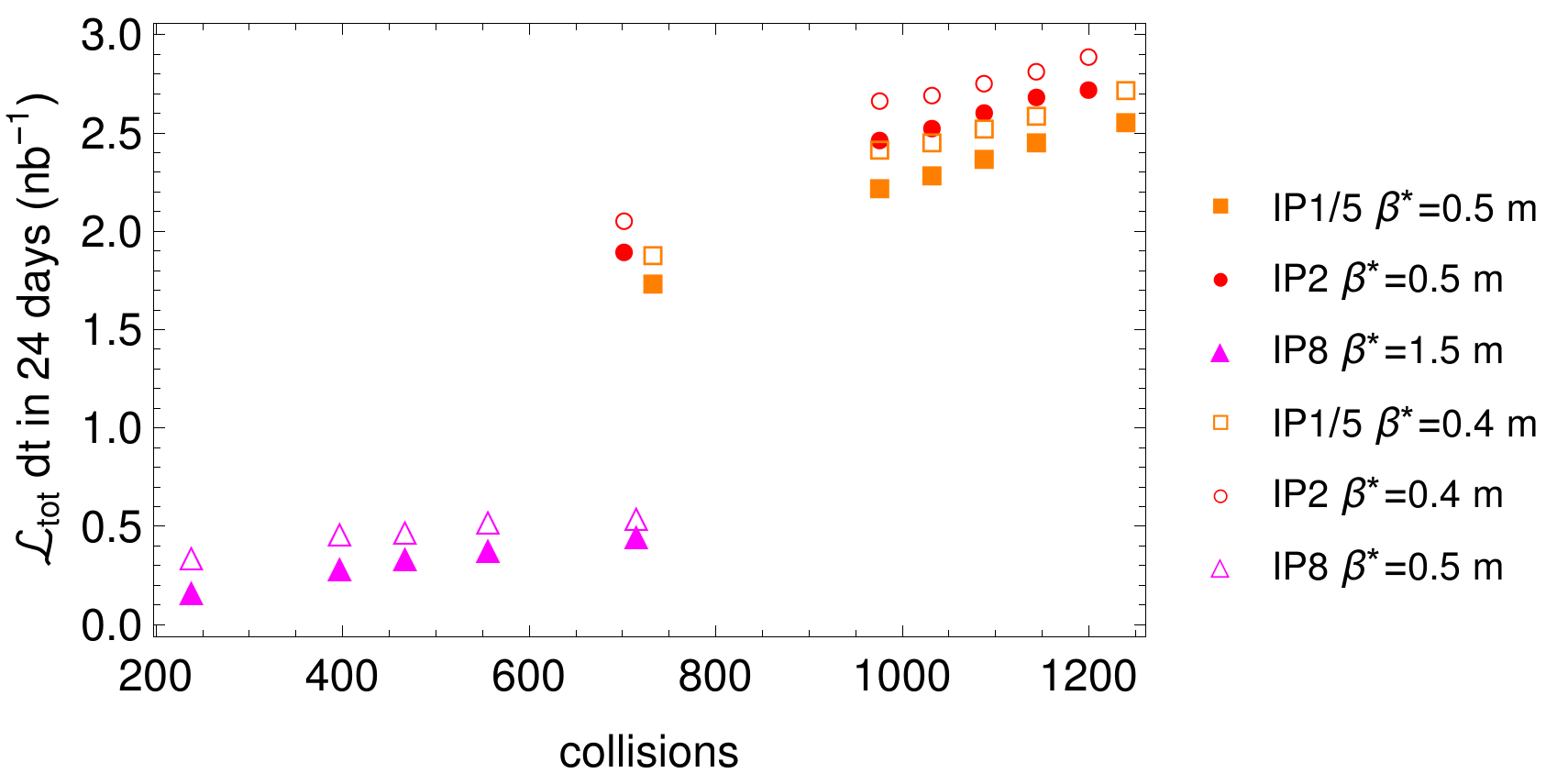}
    \caption{ Calculated integrated luminosity during a one-month Pb-Pb run at each experiment, as a function of the number of colliding bunches, comparing the baseline configuration (filled markers) with a reduced-\bs configuration (open markers). 
    The luminosity has been calculated using CTE simulations and Eq.~(\ref{eq:avgLum})--(\ref{eq:Ltot}), assuming an operational efficiency of $\eta$=0.5 and  24~days available for physics operation. }
    \label{fig:lum40cm}
\end{figure}

An optimistic extrapolation of aperture measurements with protons~\cite{bruce19_evian} leads us to study with CTE a configuration with \bstarval{0.4} at IP1, IP2, and IP5, and \bstarval{0.5} at IP8, while keeping the baseline crossing angles. This relies also on the assumption that there is some margin to decrease the normalized beam-beam separation, which in the baseline is assumed to be larger than in the 2018 proton operation. The CTE results are illustrated in Fig.~\ref{fig:lum40cm}, which shows $\Ltot$  per one-month run as a function of the number of bunches colliding per IP, together with baseline performance in Table~\ref{tab:int_lum_PbPb}. In the configuration with smaller \bstar, we observe a gain in $\Ltot$ of 6--9\% at IP1, IP2, and IP5, while the gain at IP8 is up to a factor~2, depending on filling scheme. 
The gain is much smaller with more collisions at IP8, since the levelling time   approaches $\Topt$. 

We have also investigated other performance enhancements for comparison. 
If the IP1/5 net half crossing angle could be reduced to \qty{100}{\murad} as at IP2, a gain of 6--7\% at IP1/5 and a loss of 2--4\% at IP2 are predicted. 
The crossing angle has a larger effect at IP8, where a net half angle of \qty{-100}{\murad} would bring a gain in $\Ltot$ of around 13\%. 
It could also be envisaged not to use luminosity levelling at IP1/5, since the detectors are not limited and there is still some margin to the BFPP quench limit. However, using the above assumptions, the simulated gain in $\Ltot$ is   limited at about 1--2\%, with a similar loss at IP2/8. In this case, $\Topt$ is significantly different between IP1/5 and IP2. Shorter fills with a $\Topt$ calculated for IP1/5 could be compensated with a filling scheme that redistributes some collisions to IP2. 
Such a scheme could potentially provide gains for all experiments, but a detailed study is left as future work.  

All these paths  to increased $\Ltot$   still need to be demonstrated through further studies.

\subsection{p-Pb performance in the HL-LHC baseline configuration}
\label{sec:pPbHL_perf} 

\begin{figure*}[!tbh]
    \centering
    \includegraphics[width=\textwidth]{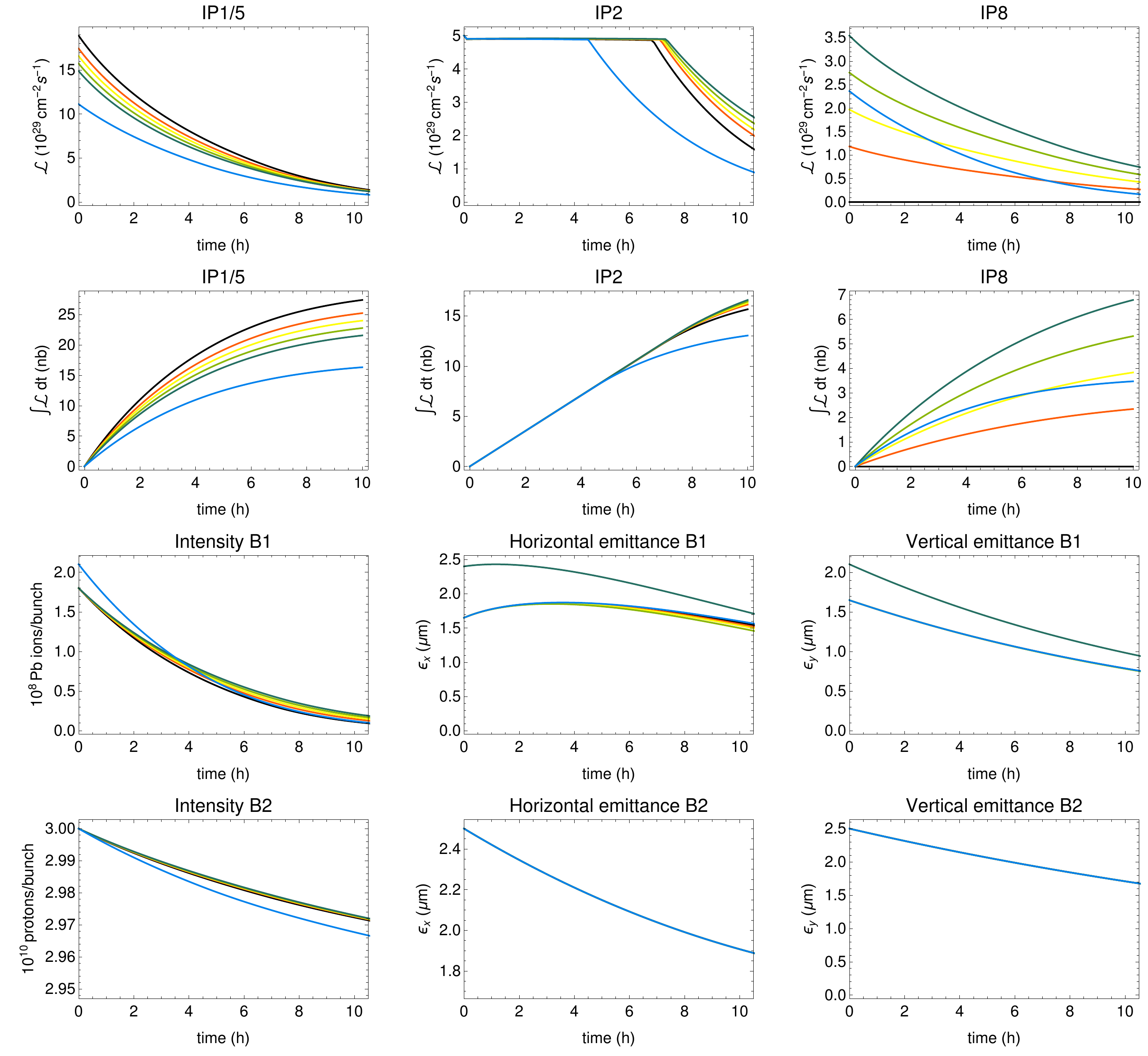}
    \includegraphics[trim={0 0 0 6.3cm},clip,width=0.6\textwidth]{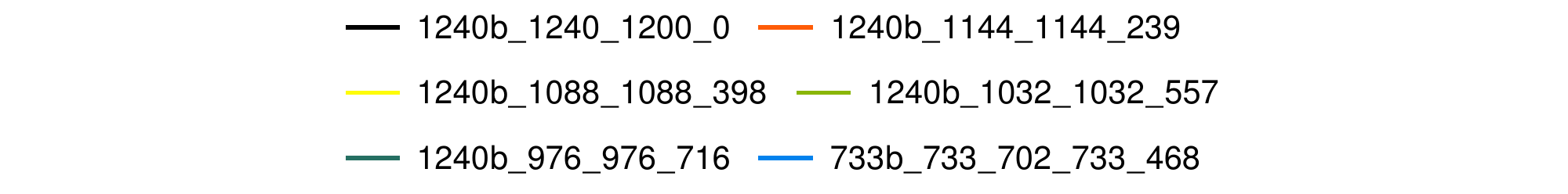}
    \caption{ The simulated HL-LHC p-Pb performance from   MBS in terms of instantaneous luminosity (1st row) and integrated luminosity (2nd row) during a typical fill for the considered filling schemes from Ref.~\cite{bruce20_HL_ion_report}, shown together with the evolution of the intensity and normalized emittances of the Pb ions in B1 (3rd row) and protons in B2 (4th row). }
    \label{fig:HL-sim-pPb}
\end{figure*}

For p-Pb operation, we base our running scenario on the detailed considerations in Refs.~\cite{marc-thesis,bruce20_HL_ion_report}. 
We assume the same filling patterns as for Pb-Pb. This is an approximation, since the injection sequence for protons is different to that for Pb. While proton injection is expected to be rather flexible we account for the possibility that it cannot  perfectly match the Pb filling pattern, by subtracting 5\% of the total integrated luminosity  as in Ref.~\cite{marc-thesis}.

In the simulations, B1 is assumed to contain Pb and B2 protons, but interchanging the beams does not influence the outcome at this level. 
We assume the same future Pb beam parameters as for Pb-Pb (see Table~\ref{tab:lhc}) and that the proton beam has intensities chosen to be at \expfor{3}{10} protons per bunch and normalized emittances around \mum{2.5}. 

We include in MBS a phenonemological 100~h non-collisional lifetime for the Pb beam as in Sec.~\ref{sec:PbPbHL_perf}.
In addition, as explained in~\cite{marc-thesis} and based on fits to 2016 p-Pb data, an IP-dependent lifetime of 38~h is included per collision for bunches colliding in IP1 and IP5, 48~h  in IP2 and 317~h in IP8.
For the proton beam, a 5842~h non-collisional lifetime is used for all bunches, as observed in 2016~\cite{marc-thesis}. It is assumed that IP2 is levelled at $\Lu=$\lumTN{5},   a factor~5 higher than in 2016, 
thanks to the   ALICE upgrade~\cite{ALICE_LOI_2014,Abelev_et_al_2014}. 
As in 2016, IP1, IP5, and IP8 are not levelled. 
The assumed burn-off cross sections for p-Pb are shown in Table~\ref{tab:cross_sections}. 
Because of their strong dependence on $Z$, the electromagnetic processes are almost negligible in comparison to Pb-Pb operation.

\begin{table*}[]
\centering
\begin{tabular}{lccc} \hline
Filling scheme             & $\Ltot$  IP1/5   & $\Ltot$ IP2             & $\Ltot$ IP8  \\ \hline 
\fs{1240}{1240}{1200}{0}   & 677 [705]                & 306 [313]                           & 0 [0]                            \\ 
\fs{1240}{1144}{1144}{239}  & 634 [647]                & 309 [316]                           & 45 [52]                         \\ 
\fs{1240}{1088}{1088}{398} & 605 [613]                & 308 [317]                           & 73 [85]                         \\ 
\fs{1240}{1032}{1032}{557} & 583 [580]                & 311 [319]                           & 103 [119]                         \\ 
\fs{1240}{976}{976}{716} & 558 [547]                & 312 [320]                           & 135 [152]                         \\ 
\fs{733}{733}{702}{468}   & 415 [431]                & 287 [294]                           & 86 [88]                        \\ \hline
\end{tabular}
\caption{
Integrated luminosity (in \nb) during a one-month p-Pb run at each experiment for the considered filling schemes from Ref.~\cite{bruce20_HL_ion_report}, using Eq.~(\ref{eq:avgLum})--(\ref{eq:Ltot}). The first value comes from  CTE  while the one in square brackets is from MBS. }
\label{tab:int_lum_pPb}
\end{table*}

The simulated time evolution of the luminosity and beam parameters from MBS are shown in Fig.~\ref{fig:HL-sim-pPb}. The proton intensity remains almost unchanged as the luminosity losses are small in comparison to the total intensity. 
The Pb beam loses about one order of magnitude  of its intensity in 10~h. 

The IP1/5 luminosity  starts  
from a peak of $\Lu=$\lumTN{22.3}, about twice  what was achieved in 2016,
and then decays rapidly.
IP2 needs to be levelled significantly longer than with Pb-Pb, with typical levelling times of 6--7~h. 
The optimal fill time, calculated to maximise $\Lavg$ in Eq.~(\ref{eq:avgLum}) 
and assuming $\Tta=200$~minutes, is therefore rather different between IP2 (typically around 8~h) and IP1/5 (typically around 4.8~h). 
The fill time used for the calculation of $\Ltot$ for p-Pb is therefore estimated as the geometric mean of the two optima,  typically $\Tf\approx\qty{6.2}{h}$.

The calculated $\Ltot$ per one-month run is given in Table~\ref{tab:int_lum_pPb} with the same assumptions as for Pb-Pb. 
Results from CTE and MBS agree to within 5\% at  IP1/5 and IP2. Slightly larger discrepancies af about 15\% are found for the schemes with few collisions at IP8. 

At IP1 and IP5, $\Ltot$ is in the range of 530--690~\nb per run for the 50~ns schemes, which is significantly higher than the 310~\nb obtained at IP2, mainly  because only IP2 is levelled.  
The loss in $\Ltot$ with the 75~ns backup scenario is about 20--40\% at IP1/5, but only about around 10\% for IP2, thanks  to the levelling.

At IP8, the maximum simulated $\Ltot$ in a one-month run over the   filling schemes is about 150~\nb. 
As for Pb-Pb, we could envisage improvements such as a smaller \bstar\ and smaller crossing angles if a higher $\Ltot$ will be needed.

The above estimates are for p-Pb operation as envisaged until the end of Run~4. Studies with MBS of the performance in collisions between protons and lighter nuclei, as proposed for after Run~4~\cite{YR_WG5_2018}, are presented in Ref.~\cite{Jebramcik:IPAC2019-MOPMP024}.

\section{Conclusions}
\label{sec:conclusions}

We have presented two independent simulation codes for beam and luminosity evolution in heavy-ion colliders, CTE and MBS, which are based on different underlying principles. CTE tracks bunches of macro-particles, subject to various physical effects, while MBS numerically solves a system of ordinary differential equations. MBS has the advantage of individually simulating every single bunch in the machine, while CTE can instead  track non-Gaussian distributions and has more physical effects included, such as emittance blowup from collisions and machine aperture. 

A thorough benchmark with data from 30~physics fills in the 2018 LHC Pb-Pb run shows an excellent agreement between both codes and measurements of the luminosity and intensity evolution. 
Given the starting conditions and fill duration, the integrated luminosity can typically be predicted within a few percent at the main experiments. Minor discrepancies are observed in emittances and bunch length, where the latter could be corrected in CTE by introducing a suitable coupling between the horizontal and vertical emittances. 

The codes have then been used to predict the performance in future LHC heavy-ion runs with updated beam parameters, filling schemes and operational assumptions. We estimate an integrated luminosity per one-month Pb-Pb run of about 2.2--2.6~\nb in ATLAS (IP1) and CMS (IP5), 2.4--2.8~\nb in ALICE (IP2) and up to about 0.5~\nb in LHCb (IP8), assuming a 50\% operational efficiency and   24 days  of physics operation after the initial commissioning. 

With p-Pb, the integrated luminosity per one-month run is estimated to lie in the range of 530--690~\nb at ATLAS and CMS, at about 310~\nb at ALICE, and up to about 150~\nb at LHCb.

A sensitivity study shows that the results are rather robust against uncertainties in the IBS model and changes in the non-collisional lifetime. On the other hand they are rather sensitive to any change in bunch intensity and operational efficiency and, to a lesser extent, to variations of the initial emittances. 
The present projections might thus be exceeded, e.g., if the operational efficiency  or  injected bunch intensity is higher. 
Other ways to gain   integrated luminosity   could include smaller \bstar, smaller crossing angles, or modified filling schemes.

\subsection*{Acknowledgements}
Research supported partly by the HL-LHC project.

We  thank T.~Argryopoulos, H. Bartosik, R. De Maria, N. Fuster-Martinez, N. Mounet,S. Redaelli, G. Rumolo E.~Shaposhnikova, and H.~Timko
         for useful discussions and collaborations, in particular on studies for the operational scenario for future Pb operation. 
We especially thank M.~Blaskiewicz for early collaboration and   
implementation of some elements used in CTE.


\bibliographystyle{spphys2}
\bibliography{all-refs2}

\end{document}